\documentclass[submitting]{nst}

\usepackage{subfigure,dcolumn}
\usepackage[T2A,T1]{fontenc}
\usepackage[russian,english]{babel}

\usepackage{listings}
\usepackage{graphicx}
\usepackage{bm}
\usepackage{multirow}
\usepackage{CJK}

\begin{document}

\begin{CJK*}{UTF8}{gbsn}
\title{Nuclear landscape based on point-coupling density functional with localized exchange terms}\thanks{This work is supported by the National Key Research and Development (R\&D) Program under Grant No. 2021YFA1601500, the National Natural Science Foundation of China under Grant No. 12075104, No. 12147101, and No. 12447106, the Fundamental Research Funds for the Central Universities lzujbky-2023-stlt01, the Institute for Basic Science funded by the Korean government (Grant No. IBS-R031-D1), the funding of the China Institute of Atomic Energy (Grants No. YC010270525794 and No. PA010271225779), the Lingchuang Research Project of China National Nuclear Corporation under Grant No. CNNC-LCKY-2024-082, and the Science and Technology Innovation Leading Talent Project of Gansu Province (Project No. 25RCKA025).}
\author{Y.\,N.\,Huang\,(黄亚男)}
\affiliation{School of Nuclear Science and Technology, Lanzhou University, Lanzhou 730000, China}
\affiliation{Frontier Science Center for Rare isotope, Lanzhou University, Lanzhou 730000, China}
\author{Q.\,Zhao\,(赵强)}
\email[Corresponding author, ]{Q. Zhao, zhaoqiang@ciae.ac.cn}
\affiliation{China Institute of Atomic Energy, P.O. Box 275, Beijing, 102413, China}
\affiliation{Center for Exotic Nuclear Studies, Institute for Basic Science, Daejeon 34126, South Korea}
\author{Y.\,F.\,Niu\,(牛一斐)}
\email[Corresponding author, ]{Y. F. Niu, niuyf@sjtu.edu.cn}
\affiliation{School of Physics and Astronomy, Shanghai Key Laboratory for Particle Physics and Cosmology, and Key Laboratory for Particle Astrophysics and Cosmology (MoE), Shanghai Jiao Tong University, Shanghai 200240, China}
\date{\today}

\begin{abstract}

%1.work
%2.problem: PSO restoration, shell structure
%3.prospective: deformation, study on heavy nuclei

Nuclear landscape is initially explored in the framework of relativistic Hartree-Bogoliubov theory under the spherical approximation, adopting the newly developed PCF-PK1 density functional. The functional effectively incorporates exchange terms via the Fierz transformation and explicitly includes the tensor coupling. We analyze the limits of the nuclear landscape and nuclear ground state properties including binding energies, charge radii, $\alpha$-decay energies, compared with other functionals and experiment. In the present calculations, 7210 nuclei are predicted to be bound with the root-mean-square deviation of binding energies 7.170 MeV. The removal of spurious shell closure at $Z=58$ and 92 is discussed by shell gaps and single-particle spectra. For superheavy nuclei, potential magic numbers beyond $^{208}$Pb are studied. The inclusion of the tensor coupling in PCF-PK1 helps restore the pseudospin symmetry, leading to a less pronounced shell closure at $Z=120$.

\keywords{Nuclear binding energies $\cdot$ Localized exchange terms $\cdot$ Spurious shells}

\end{abstract}

\maketitle
\end{CJK*}

\section{Introduction}

%%% 1. stellar nucleosynthesis, especially r-process needs theoretical calculations 2. landscape and drip line are not clear, needs reliable description 
Nuclear mass, being one of the most fundamental properties of atomic nuclei, serves as a cornerstone for understanding nucleonic interactions \cite{F.Wienholtz_2013_Nature}. It plays a pivotal role in elucidating the evolution of nuclear shell structures \cite{K.Blaum_2006_PR}, the emergence of exotic phenomena such as nuclear halos \cite{Y.Yu_2024_PRL}, the nature of isospin symmetry \cite{Y.M.Xing_2025_PRL}, and the limits of the nuclear chart \cite{J.Erler_2012_Nature}. Furthermore, nuclear masses are indispensable inputs for nuclear astrophysics, essential for simulating nucleosynthesis processes \cite{J.Meng_2016_RDFT} and tracing the origin of elements \cite{B.H.Sun_2015_FP}. To date, experiments have determined the masses of just over 2500 nuclei \cite{M.Wang_2021_CPC}, with the neutron and proton drip lines firmly established only up to Ne ($Z=10$) \cite{D.S.Ahn_2019_PRL} and Np ($Z=93$) \cite{M.Thoennessen_2004_RoPP, Z.Y.Zhang_2019_PRL}, respectively. However, the simulations of the rapid neutron-capture process (r-process) nucleosynthesis, consist of nuclear mass data for more than 6000 isotopes \cite{M.Bradley_nucnet}, a vast number of which, especially those extremely neutron-rich nuclei, are beyond current experimental reach. Therefore, a reliable description and prediction of nuclear masses demands a unified and self-consistent theoretical model.

%%% mic-mac not good enough
Theoretical efforts to explore the nuclear landscape are primarily pursued through microscopic-macroscopic (mic-mac) and microscopic models. The microscopic-macroscopic approach achieves remarkable accuracy on the order of hundreds of keV for mass descriptions \cite{Y.Aboussir_1995_ADNDT, N.Wang_2014_PLB, P.Moller_2016_ADNDT, H.F.Zhang_2014_NPA}, as demonstrated by the WS4 (298 keV) \cite{N.Wang_2014_PLB} and FRDM2012 (560 keV) \cite{P.Moller_2016_ADNDT} models. However, its predictive power for extrapolations far from the valley of stability remains limited, as the parameters are fitted by available mass data.

%%% DFT advantages -> micro: SDFT, Gogny
To provide a consistent microscopic description of various nuclear properties across the nuclear landscape, the nuclear density functional theory (DFT) is required \cite{M.Bender_2003_RMP}. Within the framework of non-relativistic Skyrme DFT, significant efforts have been devoted to improving mass accuracy and addressing uncertainties near the boundaries of the nuclear chart \cite{J.Erler_2012_Nature}. The Hartree-Fock-Bogoliubov (HFB) mass models, e.g., HFB-17 and HFB-27*, achieved mass accuracies of 581 keV \cite{S.Goriely_2009_PRL_HFB17} and 500 keV \cite{S.Goriely_2013_PRC}, respectively, utilizing effective interactions containing up to 24 adjustable parameters. The Gogny-D1M mass model, with 14 adjustable parameters, reached an accuracy of 798 keV \cite{S.Goriely_2009_PRL_D1M}.

%%% CDFT advantages, detailed research
The covariant density functional theory (CDFT) includes naturally the spin-orbit interaction \cite{M.G.Mayer_1949_PR, O.Haxel_1949_PR}, inherently explains pseudospin symmetry \cite{H.Z.Liang_2015_PR}, and consistently includes time-odd fields \cite{W.Koepf_1989_NPA}, and it has been widely adopted in mass table studies. Initial work within the relativistic mean-field (RMF) framework employed the TMA interaction to compute nuclear masses for all even-even nuclei with an accuracy of 2.71 MeV \cite{D.Hirata_1997_NPA}, considering the axial deformation but neglecting the pairing correlation. The neglected pairing correlation in this RMF calculation can lead to unreasonable nuclear deformations, which in turn induce large discrepancies in the predicted r-process abundances \cite{D.Hirata_1997_NPA}. Pairing correlation plays a crucial role in atomic nuclei, as manifested, e.g., in the odd-even mass difference \cite{A.Bohr_1958_PR}. Such correlation can be taken into account within the Bardeen-Cooper-Schrieffer (BCS) theory. With the subsequent inclusion of BCS pairing, the functional NL3 was then applied in the calculations for even-even nuclei with an accuracy of 2.6 MeV \cite{G.A.Lalazissis_1999_ADNDT}. The first complete mass table based on the RMF+BCS model was constructed with the TMA functional. This table, including even-even, odd-$A$, and odd-odd nuclei, incorporates the deformation effect and achieves an accuracy of 2.118 MeV \cite{L.Geng_2005_PTP}. For exotic nuclei far from $\beta$-stability valley, the Bogoliubov transformation provides a proper treatment of continuum effects. Within the relativistic Hartree-Bogoliubov (RHB) framework, the PC-L3R interaction was used to predict nuclear masses for 7373 nuclei under the spherical approximation \cite{Z.X.Liu_2023_PLB}. Calculations for even-even nuclei using functionals like NL3*, DD-ME2, DD-ME$\delta$, and DD-PC1, which include axial deformation, resulted in accuracies smaller than 3 MeV \cite{A.V.Afanasjev_2013_PLB, S.E.Agbemava_2014_PRC}. Further inclusion of the triaxial deformation and dynamic correction energies with PC-PK1 functional improved the accuracy for even-even nuclei to 1.31 MeV \cite{Y.L.Yang_2021_PRC}. To properly treat weakly bound systems by incorporating continuum effects, the relativistic continuum Hartree-Bogoliubov (RCHB) theory \cite{J.Meng_1998_NPA} was developed for spherical nuclei, and the deformed relativistic Hartree-Bogoliubov in continuum (DRHBc) theory \cite{S.G.Zhou_2010_PRC} was developed for deformed nuclei. They solve the RHB equations either in the coordinate space \cite{J.Meng_1996_PRL, J.Meng_1998_NPA} or in a Dirac Woods-Saxon basis \cite{S.G.Zhou_2003_PRC, S.G.Zhou_2010_PRC}, thereby providing correct asymptotic behaviors for the wave functions. Using the PC-PK1 functional, mass tables were first constructed under the spherical approximation and later with quadrupole deformations under such frameworks. This led to a remarkable improvement in accuracy, from 7.960 MeV for all nuclei \cite{X.W.Xia_2018_ADNDT} to 1.518 MeV for even-even nuclei \cite{K.Y.Zhang_2022_ADNDT} and 1.433 MeV for even-$Z$ nuclei \cite{P.Guo_2024_ADNDT}.

%%% 1. spurious shell 2. Fock terms 3. tensor coupling (nonono); no fock terms -> problem -> PCF-PK1, solution -> mass table
In most applications of CDFT, the contribution of Fock terms is neglected. The relativistic Hartree-Fock (RHF) framework includes the exchange of $\pi$ mesons  \cite{W.H.Long_2008_EPL} and the $\rho$ tensor coupling \cite{W.H.Long_2007_PRC}, thereby naturally incorporating the tensor force. These contributions improve the description of single-particle energies and shell structure evolution and, in principle, require no additional parameters for spin-isospin excitations \cite{H.Z.Liang_2008_PRL, H.Z.Liang_2012_PRC2}. However, RHF calculations incur highly computational costs. In relativistic point-coupling models, the exchange terms of four-fermion terms can be easily expressed as direct ones by Fierz transformation without numerical complexities \cite{H.Z.Liang_2012_PRC1}. This feature motivated the development of PCF-PK1 functional. PCF-PK1 functional not only overcomes the computational cost associated with including exchange terms, but also successfully eliminates the spurious shell closures at $Z=58$ and $92$ that are not observed in the experiment. Its success in removing spurious shell closures stems from the inclusion of tensor coupling and the modified density dependence of the coupling strength $g_{\sigma}$ and $g_{\omega}$ \cite{J.Geng_2019_PRC, B.Wei_2020_CPC}. It is therefore essential to establish a nuclear mass table within the novel PCF-PK1 functional. As a first step, we perform systematic calculations for all nuclei with $8 \leq Z \leq 120$ under the spherical approximation.

The paper is organized as follows: The theoretical framework and numerical details are introduced in Secs. \ref{sec-2} and \ref{sec-3}, respectively. To evaluate the descriptive and predictive power, the ground state properties will be analyzed in Sec. \ref{sec-4}. Finally, A brief summary is given in Sec. \ref{sec-5}.

\section{Theoretical framework}\label{sec-2}

Following the point-coupling interaction in Ref. \cite{Q.Zhao_2022_PRC}, one starts from the Lagrangian density

\begin{widetext}
\begin{equation}\label{eq-1}
\begin{split}
    \mathcal{L} = & \bar{\psi}(i\gamma_{\mu}\partial^{\mu} - M)\psi -\frac12 \left[ \alpha_S(\bar{\psi}\psi)(\bar{\psi}\psi) + \alpha_{tS}(\bar{\psi}\vec{\tau}\psi)(\bar{\psi}\vec{\tau}\psi) + \alpha_V(\bar{\psi}\gamma_{\mu}\psi)(\bar{\psi}\gamma^{\mu}\psi) + \alpha_{tV}(\bar{\psi}\gamma_{\mu}\vec{\tau}\psi)(\bar{\psi}\gamma^{\mu}\vec{\tau}\psi) \right.\\
    & + \alpha_T(\bar{\psi}\sigma_{\mu\nu}\psi)(\bar{\psi}\sigma^{\mu\nu}\psi) + \alpha_{tT}(\bar{\psi}\sigma_{\mu\nu}\vec{\tau}\psi)(\bar{\psi}\sigma^{\mu\nu}\vec{\tau}\psi) + \alpha_{PS}(\bar{\psi}\gamma_5\psi)(\bar{\psi}\gamma_5\psi) + \alpha_{tPS}(\bar{\psi}\gamma_5\vec{\tau}\psi)(\bar{\psi}\gamma_5\vec{\tau}\psi) \\
    & + \left. \alpha_{PV}(\bar{\psi}\gamma_5\gamma_{\mu}\psi)(\bar{\psi}\gamma_5\gamma^{\mu}\psi) + \alpha_{tPV}(\bar{\psi}\gamma_5\gamma_{\mu}\vec{\tau}\psi)(\bar{\psi}\gamma_5\gamma^{\mu}\vec{\tau}\psi) \right] \\
    & - \frac12 \delta_S\partial_{\mu}(\bar{\psi}\psi)\partial^{\mu}(\bar{\psi}\psi) -e\frac{1-\tau_3}{2} \bar{\psi}\gamma_{\mu}\psi A^{\mu} - \frac14 F_{\mu\nu}F^{\mu\nu} \, ,
\end{split}
\end{equation}
\end{widetext}

\noindent where $M$ is the nucleon mass, and $\psi$ represents the nucleon field. $\vec{\tau}$ is the isospin Pauli matrix, and $A_{\mu}$ and $F_{\mu\nu}$ indicate four-vector potential and strength tensor of the electromagnetic field, respectively. Here, $\alpha_i$ stands for the coupling constant for four-fermion term, including scalar ($S$), vector ($V$), tensor ($T$), pseudoscalar ($PS$), and pseudovector ($PV$) nucleon fields and subscript ``$t$" stands for isovector fields. $\delta_S$ refers to the coupling constant of the derivative term. With no-sea approximation, the nucleon field operator is expanded as single-particle operators $\{c_{\alpha}, c_{\alpha}^{\dagger}\}$ defined by a complete set of Dirac spinors $\{\phi_{\alpha}(\bm{r})\}$,
\begin{equation}\label{eq-2}
    \psi(\bm{r}) = \sum_{\alpha}\phi_{\alpha}(\bm{r})c_{\alpha} \, , \quad \psi^{\dagger}(\bm{r}) = \sum_{\alpha}\phi_{\alpha}^{\dagger}(\bm{r})c_{\alpha}^{\dagger} \, .
\end{equation}

The RHB ground state $\Phi$ is constructed as 
\begin{equation}\label{eq-3}
    |\Phi\rangle = \prod_{k} \beta_k |0\rangle \, ,
\end{equation}

\noindent where $|0\rangle$ is the bare vacuum, $\beta_k$ denotes the quasiparticle annihilation operator, and $k$ is the quasiparticle index. $\beta_k^{\dagger}$ and $\beta_k$ are defined via the unitary Bogoliubov transformation from the single-particle operators $c_l^{\dagger}$ and $c_l$ of a complete orthogonal basis,
\begin{equation}\label{eq-4}
    \beta_{k}^{\dagger} = \sum_l\left( U_{lk}c_l^{\dagger} + V_{lk}c_l \right) \, ,
\end{equation} 

\noindent where the coefficients $U_{lk}$ and $V_{lk}$ are quasiparticle wave functions. The energy density functional (EDF) can be given by the expectation value of the Hamiltonian $H$ with respect to the RHB wave function,
\begin{equation}\label{eq-5}
    E_{\rm EDF} =  \langle \Phi|H|\Phi\rangle = E_{\rm kin} + E_{\rm 4f} + E_{\rm der} + E_{\rm em} \, .
\end{equation}

The kinetic energy can be written as
\begin{equation}\label{eq-6}
    E_{\rm kin} = \int \mathrm{d}\bm{r}\sum_{\alpha} \bar{\varphi}_{\alpha}(\bm{r})( -i\bm{\gamma}\cdot\bm{\nabla} + M )\varphi_{\alpha}(\bm{r}) \, .
\end{equation}

For the four-fermion part, both Hartree and Fock terms are considered,
\begin{equation}\label{eq-7}
\begin{split}
    & E_{\rm 4f} = E_{\rm H} + E_{\rm F} = \\
    & \; \frac12 \int {\rm d}^3\bm{r} \sum_{i,\alpha\beta} \alpha_i^{\rm HF} \left[ \bar{\varphi}_{\alpha}(\bm{r})(\mathcal{O}\Gamma)_i\varphi_{\alpha}(\bm{r}) \right] \left[ \bar{\varphi}_{\beta}(\bm{r})(\mathcal{O}\Gamma)^i\varphi_{\beta}(\bm{r}) \right] \\
    & -\frac12 \int {\rm d}^3\bm{r} \sum_{i,\alpha\beta} \alpha_i^{\rm HF}  \left[ \bar{\varphi}_{\alpha}(\bm{r})(\mathcal{O}\Gamma)_i\varphi_{\beta}(\bm{r}) \right] \left[ \bar{\varphi}_{\beta}(\bm{r})(\mathcal{O}\Gamma)^i\varphi_{\alpha}(\bm{r}) \right] \, .
\end{split}
\end{equation}

Here $\mathcal{O}\in\{ 1,\vec{\tau} \}, \Gamma\in\{ 1, \gamma_{\mu}, \gamma_5, \gamma_5\gamma_{\mu}, \sigma_{\mu\nu} \}$. $\alpha_i^{\rm HF}$ denotes the coupling constant under the Hartree-Fock approximation, where $i$ runs over all possible channels. The exchange terms can be expressed as the superposition of the direct terms via the Fierz transformation \cite{A.Sulaksono_2003_AP, W.Greiner_2009_Fierz},
\begin{equation}\label{eq-7+1}
\begin{split}
    & [ \bar{\varphi}_{\alpha}(\mathcal{O}\Gamma)_i \varphi_{\beta} ] [ \bar{\varphi}_{\beta}(\mathcal{O}\Gamma)^i \varphi_{\alpha} ] \\
    & \; =\sum_j \Lambda_{ij} [ \bar{\varphi}_{\alpha}(\mathcal{O}\Gamma)_j \varphi_{\alpha} ] [ \bar{\varphi}_{\beta}(\mathcal{O}\Gamma)^j \varphi_{\beta} ] \, ,
\end{split}
\end{equation}

\noindent with $\Lambda$ being the Fierz transformation matrix. Consequently, $E_{\rm 4f}$ can be rewritten in a direct form, 
\begin{equation}\label{eq-8}
\begin{split}
    & E_{\rm 4f} = \\
    & \; \frac12 \int \!\! \mathrm{d}\bm{r} \!\! \sum_{i,\alpha\beta} \alpha_i [ \bar{\varphi}_{\alpha}(\bm{r})(\mathcal{O}\Gamma)_i\varphi_{\alpha}(\bm{r}) ][ \bar{\varphi}_{\beta}(\bm{r})(\mathcal{O}\Gamma)^i\varphi_{\beta}(\bm{r}) ] \, ,
\end{split}
\end{equation}

\noindent where $\alpha_i$ is defined as
\begin{equation}\label{eq-9}
\begin{split}
    \alpha_i &= \sum_j C_{ij} \alpha_j^{\rm HF}, \\
    i,j &= \{ S, tS, V, tV, T, tT, PS, tPS, PV, tPV \} \, .
\end{split}
\end{equation}

See detailed form of matrix $C=1-\Lambda^T$ in Ref. \cite{Q.Zhao_2022_PRC}. The coupling constants $\alpha_i(i=S, tS, V, tV)$ are in density dependent form and they follow
\begin{equation}\label{eq-10}
    f_i(x) = a_i\frac{1+b_i(x+d_i)^2}{1+c_i(x+d_i)^2} \, ,
\end{equation}

\noindent where $\rho_{\rm sat.}$ is nucleon saturation density and $x=\rho/\rho_{\rm sat.}$. Note the matrix $C$ has rank 5, indicating only five independent parameters. With the coupling constants for the $S, tS, V, tV$ channels treated as free parameters, the remaining channels read
\begin{align}\label{eq-10+1}
    & \alpha_{tT} = ( -\alpha_S + 3\alpha_{tS} + \alpha_{V} - 6\alpha_{tV} + 6\alpha_T )/18 , \\ 
    & \alpha_{PS} = ( -\alpha_S - 6\alpha_{tS} - 4\alpha_{V} + 12\alpha_{tV} - 12\alpha_T )/3 \, , \\
    & \alpha_{tPS} = ( -4\alpha_S + 3\alpha_{tS} + 8\alpha_{V} - 24\alpha_{tV} - 12\alpha_T )/9 , \\
    & \alpha_{PV} = ( 2\alpha_S + 3\alpha_{tS} + 2\alpha_{V} + 3\alpha_{tV} + 6\alpha_T )/3 \, , \\
    & \alpha_{tPV} = ( 2\alpha_S + 3\alpha_{tS} + 5\alpha_{V} - 6\alpha_{tV} + 6\alpha_T )/9 \, .
\end{align}

The derivative term, $E_{\rm der}$ and electromagnetic part, $E_{\rm em}$ are respectively written as
\begin{equation}\label{eq-11}
    E_{\rm der} = - \frac12 \delta_S \!\! \int\mathrm{d}\bm{r}\sum_{\alpha\beta}\bm{\nabla}[\bar{\varphi}_{\alpha}(\bm{r})\varphi_{\alpha}(\bm{r})] \bm{\nabla}[\bar{\varphi}_{\beta}(\bm{r})\varphi_{\beta}(\bm{r})] \, ,
\end{equation}

\noindent and
\begin{equation}\label{eq-12}
\begin{split}
    E_{\rm em} = & \frac12 \iint\mathrm{d}\bm{r}\mathrm{d}\bm{r}' \sum_{\alpha\beta} \frac{e^2}{4\pi}\left[ \bar{\varphi}_{\alpha}(\bm{r})\gamma_{\mu}\frac{1-\tau_3}{2}\varphi_{\alpha}(\bm{r}) \right] \\
    & \; \frac{1}{|\bm{r}-\bm{r}'|} \left[ \bar{\varphi}_{\beta}(\bm{r}')\gamma^{\mu}\frac{1-\tau_3}{2}\varphi_{\beta}(\bm{r}') \right] \, .
\end{split}
\end{equation}

The pairing term reads
\begin{equation}\label{eq-13}
    E_{\rm pair} = \frac14 \sum_{\mu\nu\mu'\nu'} \kappa_{\mu\mu'}^* \langle\mu\mu'|V^{pp}|\nu\nu'\rangle \kappa_{\nu\nu'} \, ,
\end{equation}

\noindent where $\kappa_{\mu\mu'}=\langle \Phi|c_{\mu'}c_{\mu}|\Phi\rangle$ is pairing tensor. $V^{pp}$ in Eq. \eqref{eq-13} follows a separable form of the Gogny force \cite{Y.Tian_2009_PLB},
\begin{equation}\label{eq-14}
    V^{pp}(\bm{r}_1,\bm{r}_2,\bm{r}_1',\bm{r}_2') = - G\delta(\bm{R} - \bm{R}')P(\bm{r})P(\bm{r}')\frac12(1 - P^{\sigma}) \, ,
\end{equation}

\noindent where $\bm{R}=(\bm{r}_1 + \bm{r}_2)/2, \bm{r} = (\bm{r}_1 - \bm{r}_2)$. $P(\bm{r})$ has a Gaussian expression,
\begin{equation}\label{eq-15}
    P(\bm{r}) = \frac{1}{(4\pi a^2)^{3/2}} e^{-r^2/4a^2} \, .
\end{equation}

The relativistic Hartree-Fock-Bogoliubov (RHFB) equation can be derived by the variational procedure,
\begin{equation}\label{eq-16}
    \begin{pmatrix}
        \hat{h}_D - \lambda & \hat{\Delta} \\
        -\hat{\Delta} & -\hat{h}_D^* + \lambda \\
    \end{pmatrix}\!\!
    \begin{pmatrix}
        U_{\alpha} \\
        V_{\alpha}
    \end{pmatrix} = E_{\alpha} \!\!
    \begin{pmatrix}
        U_{\alpha} \\
        V_{\alpha}
    \end{pmatrix} \, ,
\end{equation}

\noindent where $\hat{h}_D$ is the single nucleon Dirac Hamiltonian, $\lambda$ is the Fermi energy, $\Delta$ is the pairing field. $U_{\alpha}, V_{\alpha}$ and $E_{\alpha}$ denote quasiparticle wave functions and energy. The Dirac Hamiltonian reads
\begin{equation}\label{eq-17}
    \hat{h}_D = \bm{\alpha}\cdot\bm{p} + \beta(M+S) + V^0 - \bm{\alpha}\cdot\bm{V} - i\beta\bm{\alpha}\cdot\bm{T}^0 \, ,
\end{equation}

\noindent which is described by the scalar, vector, and tensor potentials,
\begin{align}\label{eq-18} 
    S & = \alpha_S\rho_S + \alpha_{tS}\tau_3\rho_{tS} + \delta_S\Delta\rho_S  \, , \\ 
    V^0 & = \alpha_V\rho_V + \alpha_{tV}\tau_3\rho_{tV} + e\frac{1-\tau_3}{2}\bm{A}^0  \, , \\
    \bm{V} & = \alpha_V\bm{j}_V + \alpha_{tV}\tau_3\bm{j}_{tV} + e\frac{1-\tau_3}{2}\bm{A}  \, , \\
    \bm{T}^0 & = 2\alpha_T\bm{j}_t^0 + 2\alpha_{tT}\tau_3\bm{j}_{tT}^0  \, .
\end{align}

The densities and currents are defined via
\begin{align}\label{eq-19} 
    \rho_S & = \sum_{\alpha} \bar{V}_{\alpha} V_{\alpha} , \quad \rho_{tS} = \sum_{\alpha} \bar{V}_{\alpha} \tau_3 V_{\alpha} \, , \\
    \rho_V & = \sum_{\alpha} \bar{V}_{\alpha} \gamma^0 V_{\alpha} , \quad \rho_{tV} = \sum_{\alpha} \bar{V}_{\alpha} \gamma^0 \tau_3 V_{\alpha}\, , \\
    \bm{j}_V & = \sum_{\alpha} \bar{V}_{\alpha} \bm{\gamma} V_{\alpha} , \quad \bm{j}_{tV} = \sum_{\alpha} \bar{V}_{\alpha} \bm{\gamma} \tau_3 V_{\alpha}\, , \\
    \bm{j}_T^0 & = \sum_{\alpha} \bar{V}_{\alpha} i \gamma^0\bm{\gamma} V_{\alpha} , \quad \bm{j}_{tT}^0 = \sum_{\alpha} \bar{V}_{\alpha} i \gamma^0\bm{\gamma} \tau_3 V_{\alpha} \, .
\end{align}

The center-of-mass (c.m.) correction energy is calculated by
\begin{equation}\label{eq-20}
    E_{\rm c.m.} = -\frac{\langle \hat{\bm{P}}_{\rm c.m.}^2 \rangle}{2MA} \, ,
\end{equation}   

\noindent in which $A$ is the mass number, and $\hat{\bm{P}}_{\rm c.m.}$ is the total momentum in the c.m. frame.

\section{Numerical Details}\label{sec-3}

As mentioned above, the functional PCF-PK1 \cite{Q.Zhao_2022_PRC} is applied for the particle-hole channel. For the particle-particle channel, we employ a finite-range separable force with the strength $G=657.5419$ MeVfm$^3$ and the width $a=0.644$ fm. The localized RHFB equation is solved in the space of harmonic-oscillator wave functions \cite{T.Niksic_2014_CPC}.

\begin{figure}[tb]
    \includegraphics[width=0.48\textwidth]{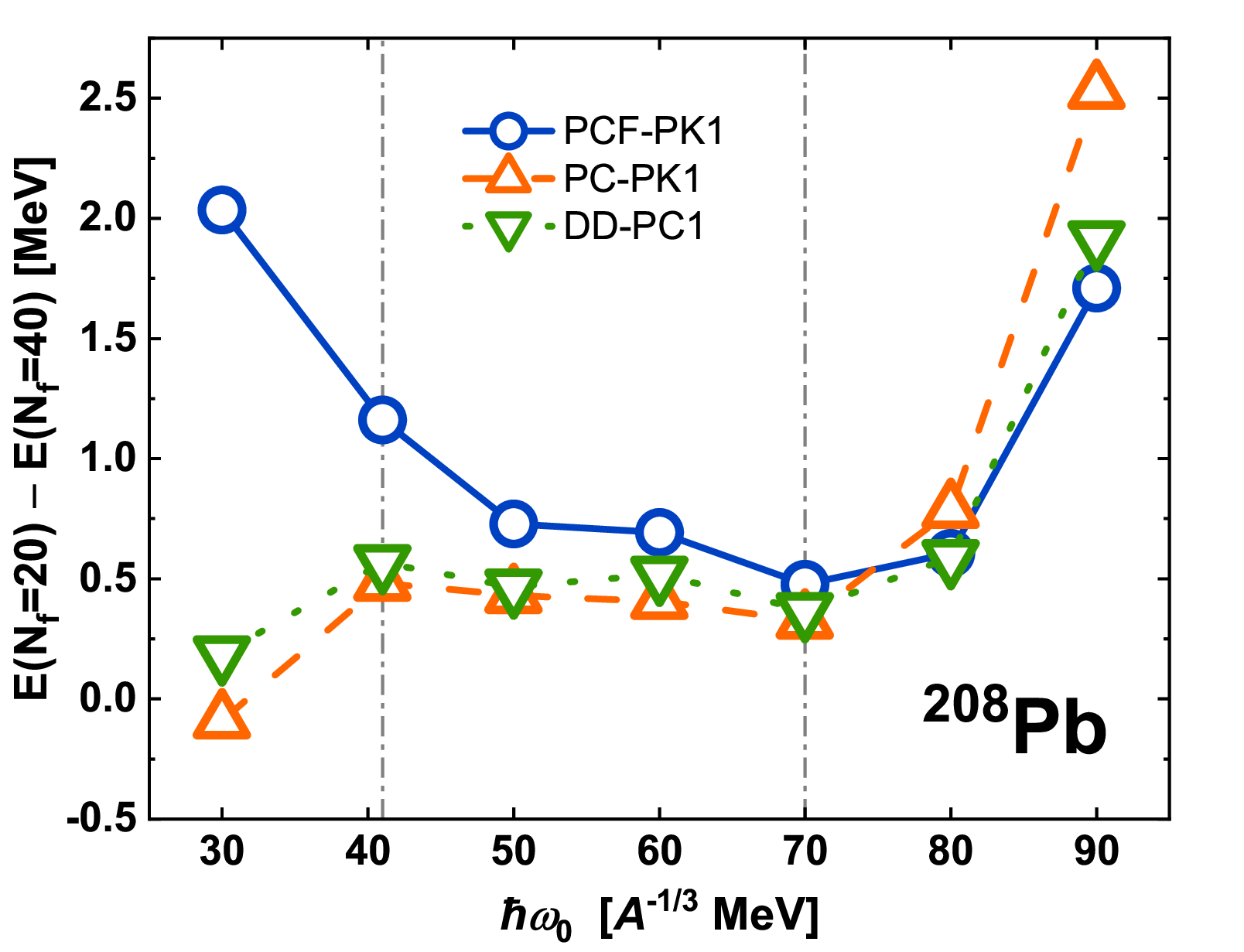}
    \caption{Total energy difference of $^{208}$Pb between calculations of $N_{\rm f}=20$ and $N_{\rm f}=40$ as a function of oscillator frequency $\hbar\omega_0$. Blue, orange, and green symbols denote results for the functionals PCF-PK1, PC-PK1 \cite{P.W.Zhao_2010_PRC}, and DD-PC1 \cite{T.Niksic_2008_PRC}, respectively. The dashed grey line indicates oscillator frequency $\hbar\omega_0=41$ and 70 $A^{-1/3}$ MeV.}
    \label{fig-1}
\end{figure}

The choice of the oscillator frequency $\hbar\omega_0$ is explained using $^{208}$Pb as an example. Fig. \ref{fig-1} depicts the nuclear total energy difference $\delta E$ between calculations with major shell numbers $N_{\rm f}=20$ and $40$. For PC-PK1 and DD-PC1, the $\delta E$ remains stable around 0.5 MeV across the range $\hbar\omega_0=41 \sim 70 A^{-1/3}$. Therefore, any value within this interval ensures the convergence, and $\hbar\omega_0=41 A^{-1/3}$ is commonly used (e.g., Refs. \cite{S.E.Agbemava_2014_PRC, Y.L.Yang_2021_PRC}). In contrast, for PCF-PK1, only $\hbar\omega_0=70 A^{-1/3}$ yields the minimal $\delta E$, which is 0.5 MeV lower than that at $\hbar\omega_0=41 A^{-1/3}$. Consequently, $\hbar\omega_0=70 A^{-1/3}$ is adopted in our systematic calculations within PCF-PK1.

\begin{figure}[tb]
    \includegraphics[width=0.48\textwidth]{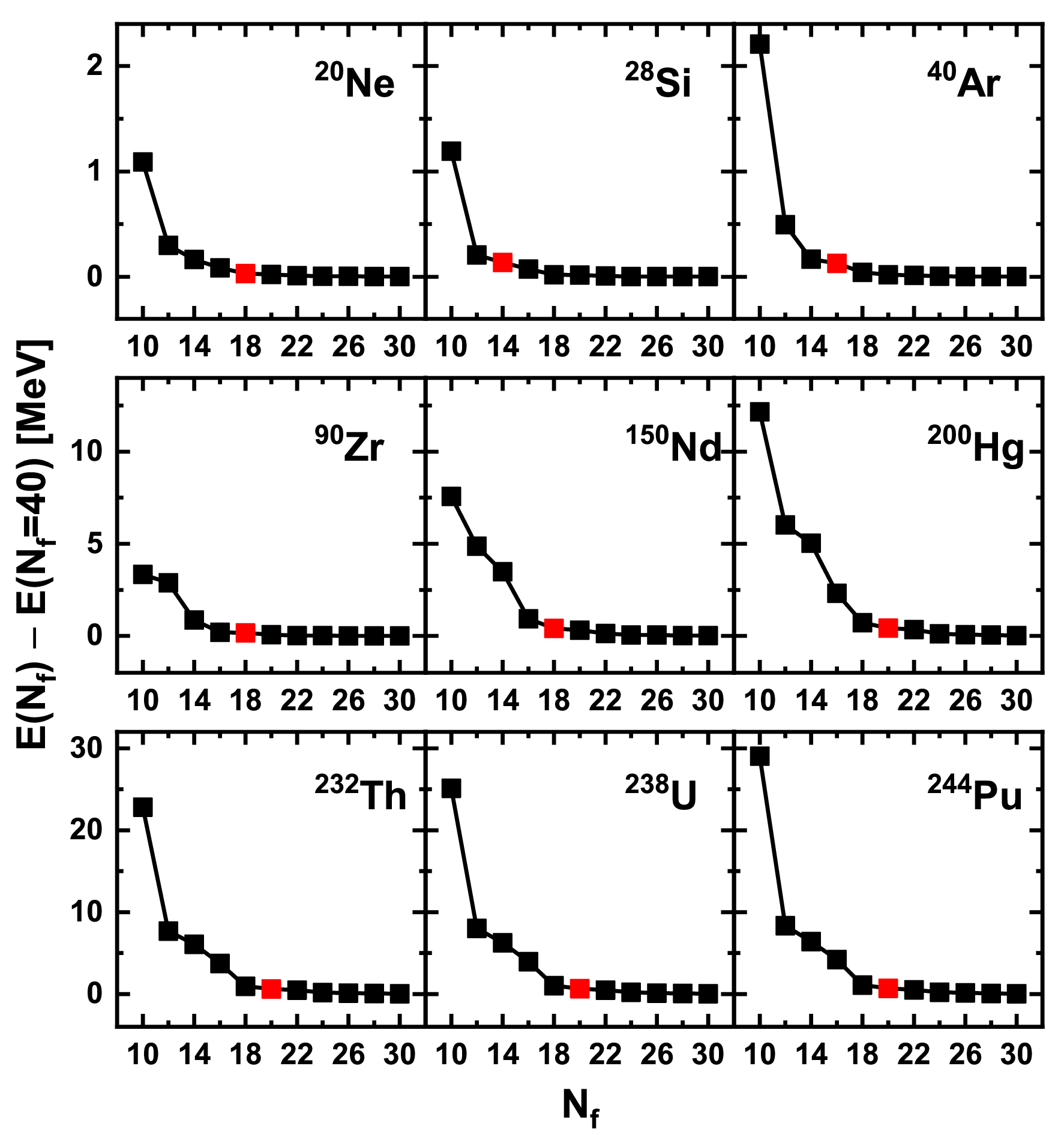}
    \caption{The total energies for nuclei  $^{20}$Ne, $^{28}$Si, $^{40}$Ar; $^{90}$Zr, $^{150}$Nd, $^{200}$Hg; $^{232}$Th, $^{238}$U, and $^{244}$Pu are plotted as a function of $N_{\rm f}$, with all values shifted relative to the results of $N_{\rm f}=40$. The red squares indicate the relative change of total energy falls within 0.05\%.}
    \label{fig-2}
\end{figure}

To determine an appropriate major shell number $N_{\rm f}$, Fig. \ref{fig-2} shows the total energies as functions of $N_{\rm f}$ for nuclei $^{20}$Ne, $^{28}$Si, $^{40}$Ar; $^{90}$Zr, $^{150}$Nd, $^{200}$Hg; $^{232}$Th, $^{238}$U, and $^{244}$Pu. All values are shifted referred to results calculated with $N_{\rm f}=40$. The red squares indicate those calculations where the relative total energy change $[E(N_{\rm f})-E(N_{\rm f}-2)]/E(N_{\rm f}-2)$ falls within the 0.05\% convergence threshold. Convergence within this threshold is achieved with 18 major shells for nuclei up to $A \leq 150$, whereas 20 major shells are required for heavier nuclei with $A \geq 200$. For simplicity, $N_{\rm f}=20$ is applied for all the nuclei in this work.

For odd-$A$ and odd-odd nuclei, the blocking effect \cite{P.Ring_many-body_1980} is taken into account. We start from an even-even nucleus with neutron number $N$ and proton number $Z$, denoted as $(N,Z)$. To obtain the ground state of the odd-$A$ nucleus with an additional neutron, $(N+1, Z)$, we perform calculations by blocking the four lowest neutron quasiparticle levels in the neighboring even-even nucleus $(N, Z)$. The one yielding the lowest total energy is determined as the ground state of $(N+1, Z)$. The procedure for an odd-$A$ nucleus with an additional proton, $(N, Z+1)$, is analogous, but the proton levels are considered. For an odd-odd nucleus $(N+1, Z+1)$, we explore all the 16 combinations of the blocked neutron and proton level from the $(N, Z)$ core, and the configuration with the lowest total energy is taken as the ground state.

\section{Results and discussions}\label{sec-4}

\subsection{Basic Properties}

\subsubsection{Nuclear Binding Energies and drip lines}

\begin{figure*}[bt]
    \includegraphics[width=0.98\textwidth]{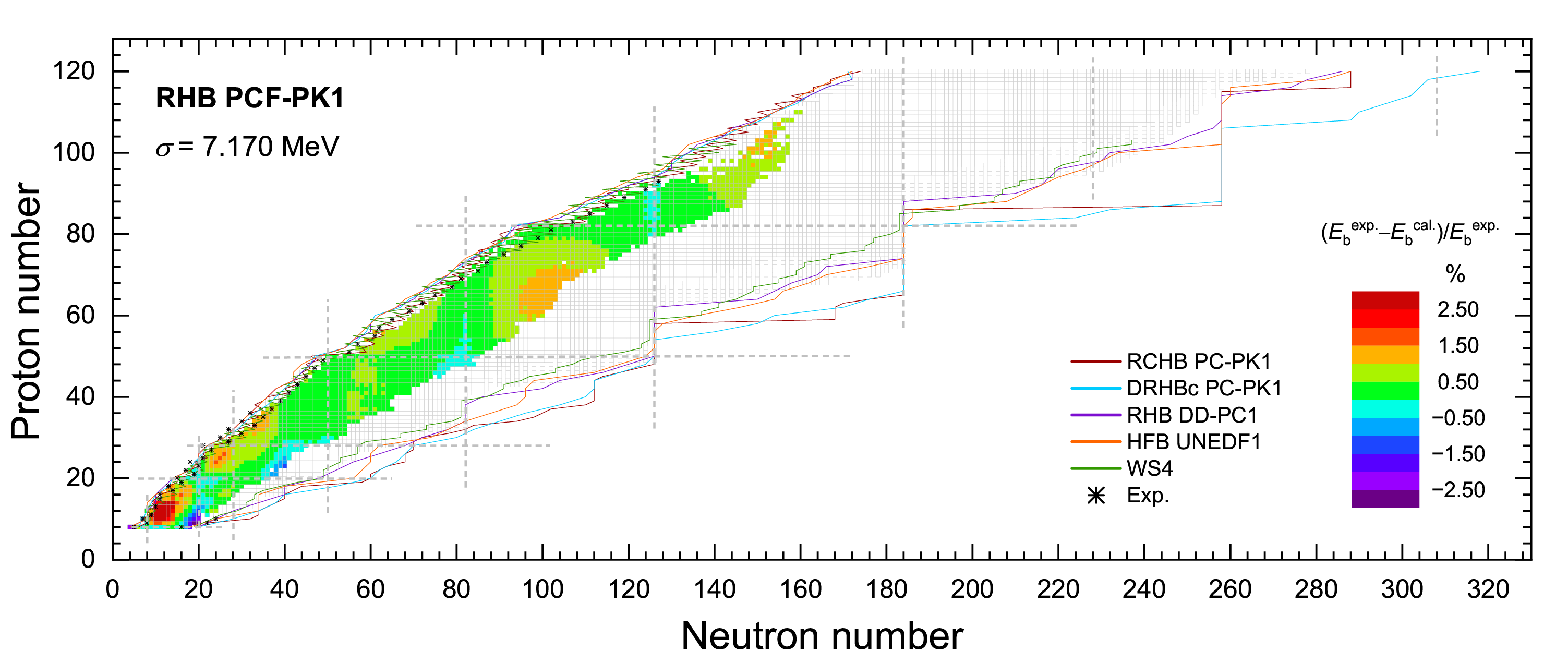}
    \caption{7210 bound nuclei ($Z=8$ to 120) are denoted by open squares. 2351 nuclei with the relative binding energy differences between calculations and the experiment \cite{M.Wang_2021_CPC} are shown by colors. Drip lines predicted by PC-PK1 \cite{X.W.Xia_2018_ADNDT, K.Y.Zhang_2022_ADNDT, P.Guo_2024_ADNDT}, DD-PC1 \cite{A.V.Afanasjev_2013_PLB, S.E.Agbemava_2014_PRC}, UNEDF1 \cite{J.Erler_2012_Nature}, and WS4 \cite{N.Wang_2014_PLB} are plotted for comparison. The drip-line nuclei confirmed in the experiment \cite{M.Thoennessen_2004_RPP, Z.Y.Zhang_2019_PRL, D.S.Ahn_2019_PRL} are shown in snow flakes.}
    \label{fig-3}
\end{figure*}

The systematic spherical calculations have been performed from $Z=8$ to 120 isotopes in the framework of RHB theory with PCF-PK1. In Fig. \ref{fig-3}, all the nuclei predicted to be bound are plotted in open squares, in which relative binding energy differences between calculated values and the experiment data $(E_{\rm b}^{\rm exp.}-E_{\rm b}^{\rm cal.})/E_{\rm b}^{\rm exp.}$ are shown by colored squares, where $E_{\rm b}$ is the nuclear binding energy. Dashed lines stand for the magic numbers $Z=8, 20, 28, 50, 82$ and $N=8, 20, 28, 50, 82, 126, 184, 228, 308$. The relative deviations near magic numbers are generally small, mostly within 0.5\%. In contrast, larger discrepancies emerge in the regions between shell closures. Systematic underestimations are particularly evident in several areas: around $(Z\sim 60, N\sim 100)$ and $(Z\sim 100, N\sim 150)$, in light nuclei with $(Z, N < 20)$ and, for proton-rich nuclei beyond $Z=20$, which can be also observed in PC-PK1 results \cite{X.W.Xia_2018_ADNDT}. These systematic deviations are likely attributable to the neglect of the deformation effect in our present spherical calculations. Near the $N=20$ and $40$ shell closures, the calculations show systematical overestimations for neutron-rich nuclei. This trend is is also found in the PC-PK1 results, indicating a common challenge for mean-field description. In Table \ref{tab-1}, the root-mean-square (rms) deviation of the calculated binding energies, $\sigma(E_{\rm b})$, from the 2351 measured data in AME2020 \cite{M.Wang_2021_CPC} is 7.170 MeV, and the deviation from the 2258 measured data in AME2012 \cite{M.Wang_2012_CPC} is 7.026 MeV. Both represent an improvement of approximately 1 MeV compared with the spherical calculations using PC-PK1 \cite{X.W.Xia_2018_ADNDT}.

In this work, we determine the drip lines using one- and two-nucleon separation energies in conjunction with the Fermi energies. For neutron drip lines, predictions for different models show large discrepancies. PC-PK1, for instance, predicts a notably more extended drip line \cite{X.W.Xia_2018_ADNDT, K.Y.Zhang_2022_ADNDT, P.Guo_2024_ADNDT}. The spherical results with PCF-PK1 are closer to the deformed predictions of DD-PC1 \cite{A.V.Afanasjev_2013_PLB, S.E.Agbemava_2014_PRC}. The underlying mechanism governing drip-line locations is complicated, however, as pointed out in Ref. \cite{R.Wang_2015_PRC}, the symmetry energy at the subsaturation density plays a vital role in the location of the neutron drip line. We find a strong negative correlation between symmetry energy at $\rho_{\rm b}=0.08$ fm$^{-3}$, $E_{\rm sym}(0.08)$, and the location of the neutron drip line. Specifically, the $E_{\rm sym}(0.08)$ for PC-PK1, PCF-PK1 and DD-PC1 are 18.71, 20.56, and 20.41 MeV, respectively. The lowest value for PC-PK1 within this trend correlates with its prediction of a more extended neutron drip line in the higher mass region. In contrast, the nearly identical neutron drip lines predicted by PCF-PK1 and DD-PC1 are consistent with their very similar values. To get a quantitative assessment, we list in Table \ref{tab-1} the rms deviations between calculated and experimental drip-line locations, defined as $\sigma_{\tau}^{\rm drip}=[ \sum_i^n (N_{i,\tau}^{\rm cal.} - N_{i,\tau}^{\rm exp.})^2 /n ]^{1/2}$. Here, $N_{i,\tau}^{\rm cal./exp.}$ stands for the neutron number of the most neutron-rich (for $\tau={\rm n}$) and proton-rich (for $\tau={\rm p}$) nuclei in the $i$-th isotopic chain according to calculation or experiment, respectively. For the O, F, and Ne isotopes, the neutron drip lines predicted by PCF-PK1 are in better agreement with experiment than those from PC-PK1. For proton drip lines, predictions from different microscopic models show little variation. This is because the proton drip line lies close to the valley of stability due to the confining effect of the Coulomb barrier.

\begin{table}[tb]
    \caption{The rms deviations of binding energies with respect to the AME2020 \cite{M.Wang_2021_CPC} and AME2012 \cite{M.Wang_2012_CPC} in unit of MeV, and $\sigma^{\rm drip}$ of neutron and proton drip lines against the experiment \cite{M.Thoennessen_2004_RPP, Z.Y.Zhang_2019_PRL, D.S.Ahn_2019_PRL}. All bound nuclei with available experimental data are included in the calculation. The results obtained with RCHB using PC-PK1 \cite{X.W.Xia_2018_ADNDT} are shown for comparison. }
    \begin{ruledtabular}
    \begin{tabular}{lcccc}
    & \multicolumn{2}{c}{RHB (PCF-PK1)} & \multicolumn{2}{c}{RCHB (PC-PK1)} \\
    \cline{2-5}
    & $\sigma(E_{\rm b})$ & Data & $\sigma(E_{\rm b})$ & Data \\
    \cline{2-5}
        AME2020 & 7.170 & 2351 & 8.103 & 2382 \\
        AME2012 & 7.026 & 2258 & 7.960 & 2284 \\
    \colrule
    & $\sigma^{\rm drip}_{\rm n}$ & $\sigma^{\rm drip}_{\rm p}$ & $\sigma^{\rm drip}_{\rm n}$ & $\sigma^{\rm drip}_{\rm p}$ \\
    \cline{2-5}
        Drip line & 3.46 & 2.02 & 5.16 & 2.42 
    \end{tabular}
    \end{ruledtabular}
    \label{tab-1}
\end{table}

\subsubsection{Charge Radii}

\begin{figure}[bt]
    \includegraphics[width=0.48\textwidth]{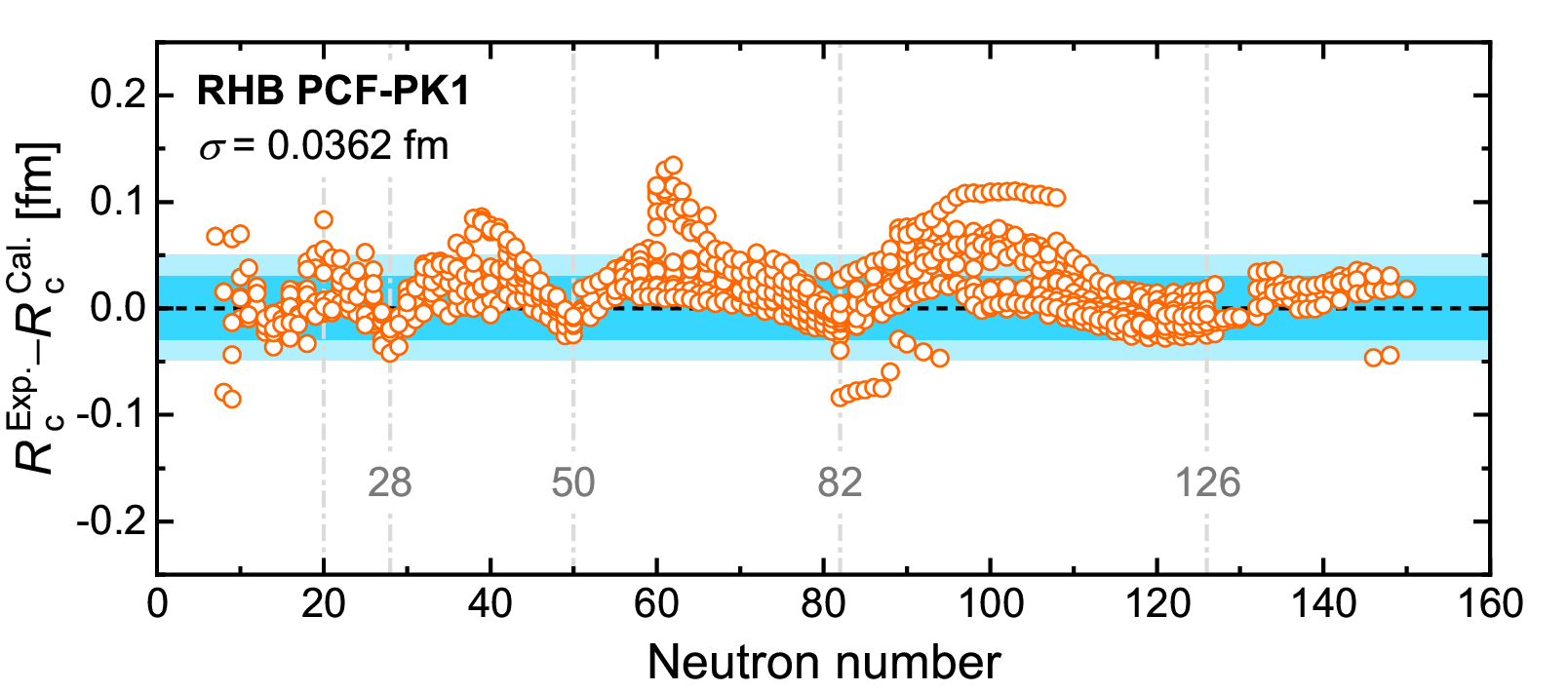}
    \caption{The rms deviations of charge radii with respect to the experiment \cite{I.Angeli_2013_ADNDT, T.Li_2021_ADNDT}. Light and deep blue regions stand for deviations are within $\pm~0.05$ fm and $\pm~0.03$ fm, respectively. The traditional magic numbers $N=20, 28, 50, 82, 126$ are shown by grey dashed lines.}
    \label{fig-4}
\end{figure}

In Fig. \ref{fig-4}, the charge radii calculated with PCF-PK1 are compared with the experiment \cite{I.Angeli_2013_ADNDT, T.Li_2021_ADNDT}. In our calculations, the deviations for 85\% of the nuclei are within $\pm 0.05$ fm, and for 70\% they are within $\pm 0.03$ fm. The $\sigma_{\rm rms}$ for all the nuclei is 0.0362 fm, which is almost the same as the result of spherical calculations within PC-PK1 \cite{X.W.Xia_2018_ADNDT}. Close to neutron traditional magic numbers, the deviations of charge radii become smaller, while in the middle of neutron shell closures they are increased up to 0.1 fm. Therefore, the underestimation of charge radii is expected to be improved if the quadrupole deformation is considered.

\subsubsection{\texorpdfstring{$\alpha$}{}-decay Energies}

There are about 1200 nuclei confirmed in the experiment \cite{M.Wang_2021_CPC} that will emit an $\alpha$ particle spontaneously, and almost all of them are situated in the mass region with $Z\geq50$. We can extract $\alpha$-decay energy $Q_{\alpha}$ as
\begin{equation}\label{eq-21}
    Q_{\alpha} (N,Z) = E_{\rm b} (N-2, Z-2) + E_{\rm b} (2, 2) - E_{\rm b} (N, Z) \, .
\end{equation}

\begin{figure}[tb]
    \includegraphics[width=0.48\textwidth]{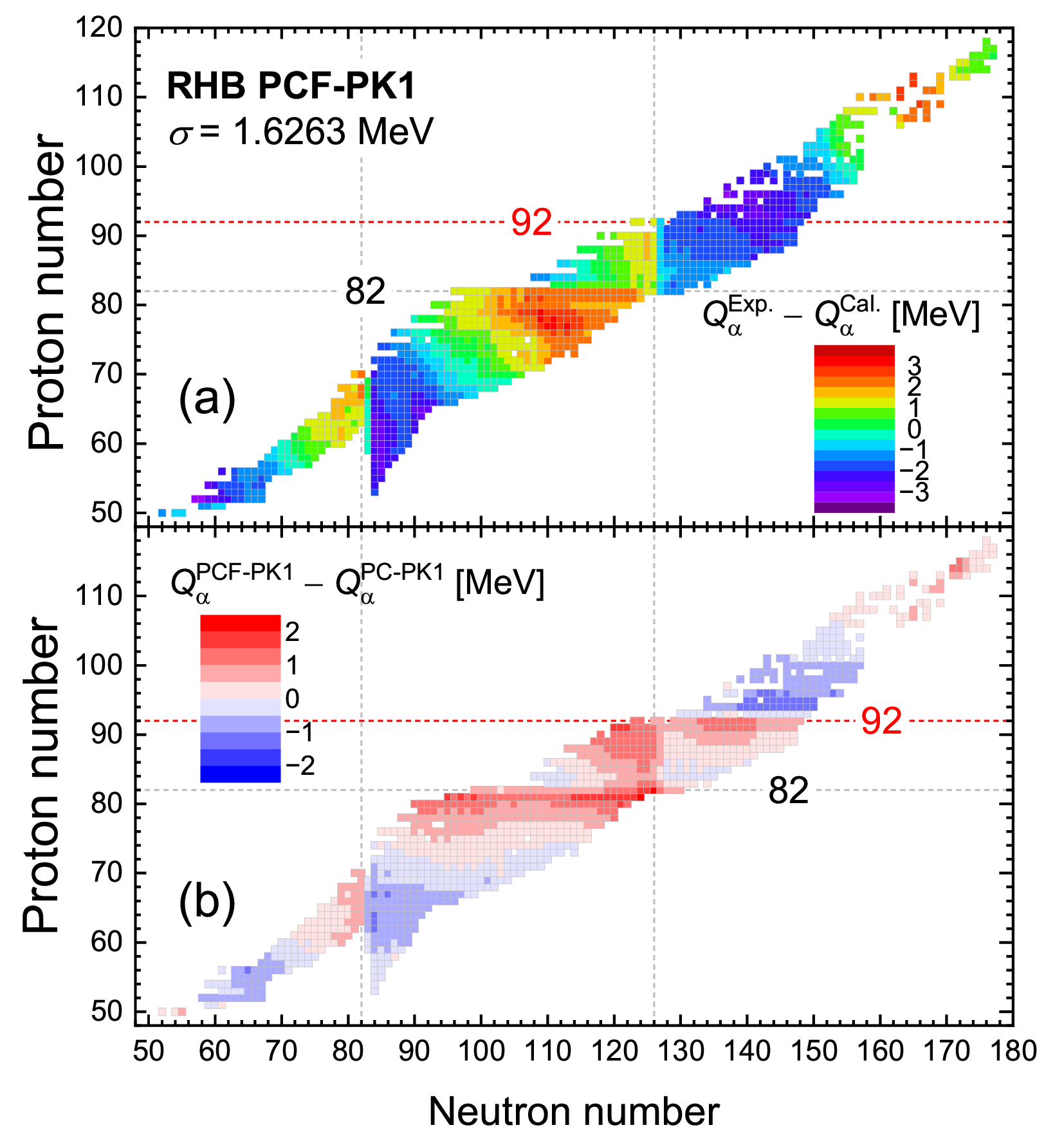}
    \caption{(a) The deviations of $Q_{\alpha}$ for PCF-PK1 against the experiment \cite{M.Wang_2021_CPC}. (b) The differences of $Q_{\alpha}$ values between PCF-PK1 and PC-PK1 results \cite{X.W.Xia_2018_ADNDT}. The traditional magic numbers $N=82, 126$ and $Z=82$ are denoted by grey dashed lines. The proton spurious shell at $Z=92$ is labelled by the red dashed line.}
    \label{fig-5}
\end{figure}
In Fig. \ref{fig-5}, the deviations of $Q_{\alpha}$ for PCF-PK1 against the experimental data are plotted in the upper panel, and the differences of $Q_{\alpha}$ values between PCF-PK1 and PC-PK1 results \cite{X.W.Xia_2018_ADNDT} are plotted in the lower panel. The rms deviation of $Q_{\alpha}$ obtained in this work is 1.6263 MeV, representing an improvement of 0.4 MeV over the PC-PK1 result. As shown in Fig. \ref{fig-5}(a), larger deviations primarily occur in the transitional regions from magic numbers toward mid-shell, particularly for nuclei with $82\leq N\leq 126$, where the nuclear deformation effect grows progressively stronger. Our calculations, constrained to the spherical symmetry, fail to capture this structural evolution, leading to the observed discrepancies. A common feature in mean-field models is the overestimation of shell closures, which leads to the overestimation of binding energy and consequently to abrupt changes in deviations at magic numbers, e.g., $Z=82$ and $N=82, 126$. Compared with the PC-PK1 results, the discrepancies around $Z=82$ and $N=82, 126$ are reduced within PCF-PK1, as shown in Fig. \ref{fig-5}(b), indicating its improved description of shell structure. In contrast to the systematic overestimation seen in Fig. \ref{fig-5}(a) for $126 \leq N \leq 150$, the abrupt reduction in $Q_{\alpha}$ values around $Z=92$ in Fig. \ref{fig-5}(b) signals an improvement in the description of the $Z=92$ shell closure within PCF-PK1.

\subsection{Shell Structures}

\subsubsection{Two-nucleon Shell Gaps}

We next examine the shell structure description of PCF-PK1 via two-nucleon shell gaps. Two-nucleon ( e.g. neutron) separation energies and shell gaps are defined as
\begin{align}\label{eq-22}
    S_{\rm 2n} (N, Z) & = E_{\rm b} (N, Z) - E_{\rm b} (N-2, Z) \, , \\
    \delta_{\rm 2n} (N, Z) & = S_{\rm 2n} (N, Z) - S_{\rm 2n} (N+2, Z) \, .
\end{align}

\begin{figure}[htb]
    \includegraphics[width=0.48\textwidth]{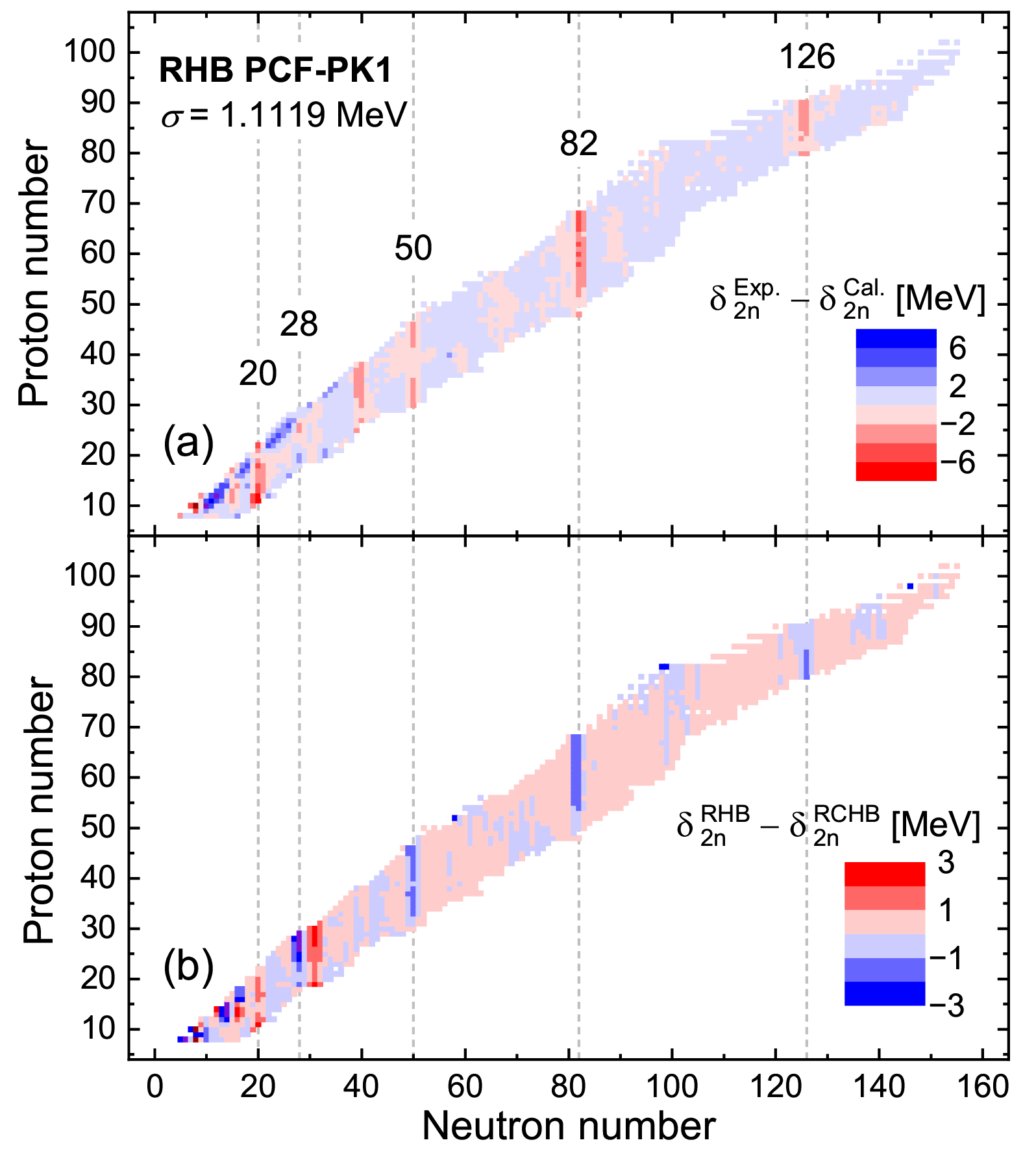}
    \caption{(a) Deviations of two-neutron shell gaps for PCF-PK1 with respect to the experiment \cite{M.Wang_2021_CPC}. (b) The differences of two-neutron shell gaps between this work and PC-PK1 results \cite{X.W.Xia_2018_ADNDT}. Grey dashed lines correspond to the traditional magic numbers $N=20, 28, 50, 82, 126$.}      
    \label{fig-6}
\end{figure}

The strength of nuclear shell closures can be quantified by two-nucleon shell gaps. We therefore begin by comparing theoretical predictions with experimental data \cite{M.Wang_2021_CPC}. Fig. \ref{fig-6}(a) shows the deviations of two-neutron shell gaps $\delta_{\rm 2n}$ for PCF-PK1 with respect to the experiment, with traditional neutron magic numbers indicated by grey vertical lines. Notably, $\delta_{\rm 2n}$ is overestimated at $N=20, 40, 50, 82, 126$, a common drawback in DFT calculations. The rms deviation of $\delta_{\rm 2n}$ for PCF-PK1 is 1.1119 MeV, which is 0.11 MeV lower than the result of PC-PK1 \cite{X.W.Xia_2018_ADNDT}. Fig. \ref{fig-6}(b) displays the difference in $\delta_{\rm 2n}$ between the results of PCF-PK1 and PC-PK1. Focusing on the magic numbers, PCF-PK1 yields smaller gaps at $N=50, 82, 126$ compared with PC-PK1, indicating a mitigation of the shell-closure overestimation. This finding is consistent with the improvement in the $Q_{\alpha}$ analysis.
\begin{figure}[hbt]
    \includegraphics[width=0.48\textwidth]{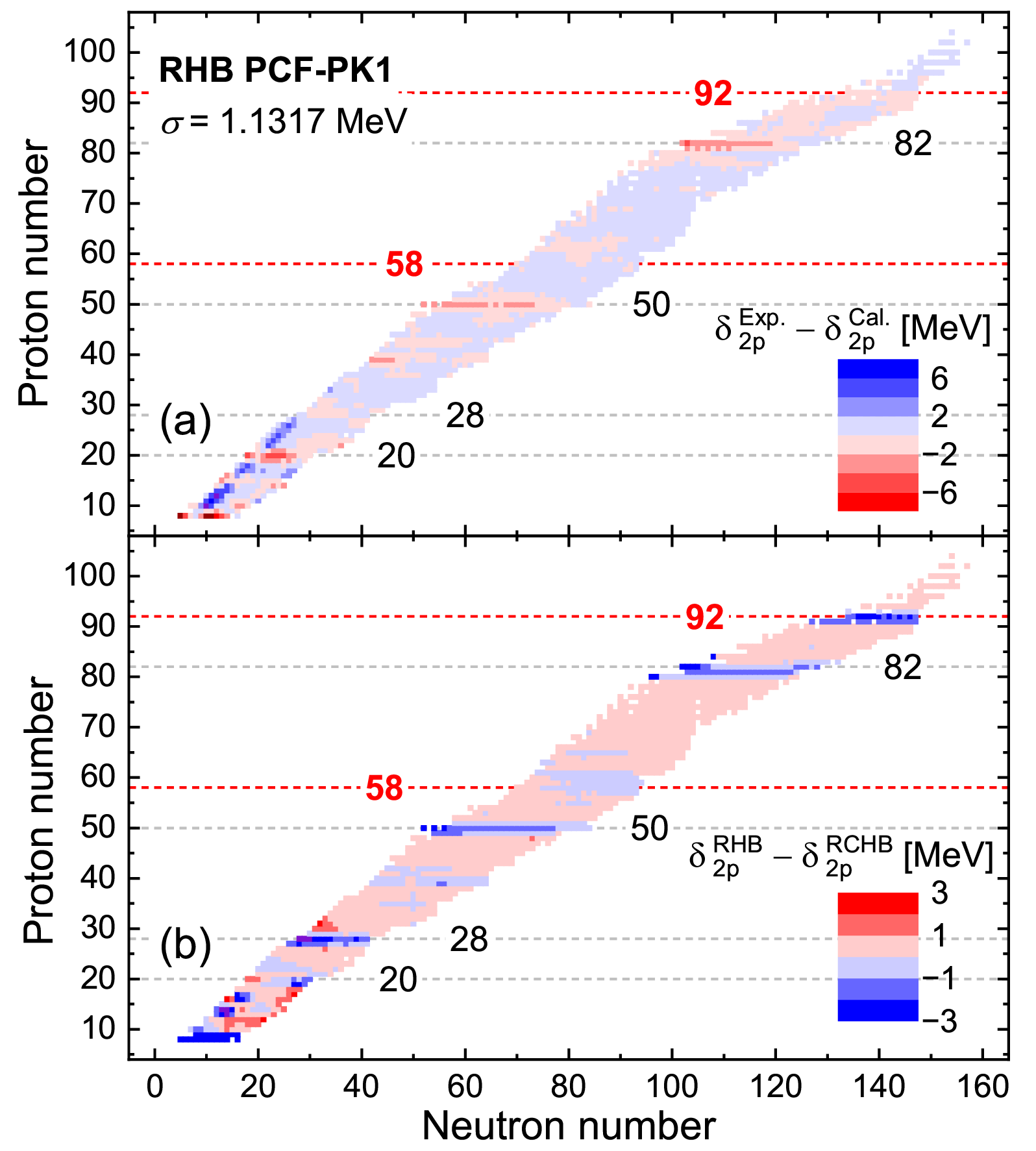}
    \caption{The same as Fig. \ref{fig-6}, but for two-proton shell gaps. Grey dashed lines correspond to the traditional magic numbers $Z=20, 28, 50, 82$ and red dashed lines stand for the spurious shells at $Z=58, 92$}
    \label{fig-7}
\end{figure}

Fig. \ref{fig-7}(a) exhibits the deviations of two-proton shell gap $\delta_{\rm 2p}$ for PCF-PK1 relative to the experiment. The rms deviation of $\delta_{\rm 2p}$ for PCF-PK1 is 1.1317 MeV, an improvement of 0.17 MeV over PC-PK1 result. Overestimations are observed at shell closures $Z=8, 20, 28, 50, 82$. As shown in Fig. \ref{fig-7}(b), PCF-PK1 reduces these overestimated gaps compared with PC-PK1 results, indicating a more reasonable shell structure. Furthermore, PCF-PK1 yields smaller $\delta_{\rm 2p}$ values at $Z=58$ and, more distinctly, at $Z=92$. For $Z=58$, while the reduction is not uniform across all isotopes, PCF-PK1 gives smaller values for most cases. This systematic reduction at the locations of known spurious shells provides clear evidence for its successful removal in the PCF-PK1 functional.

\subsubsection{Removal of Spurious Shell Closures}

\begin{figure}[htb]
    \includegraphics[width=0.48\textwidth]{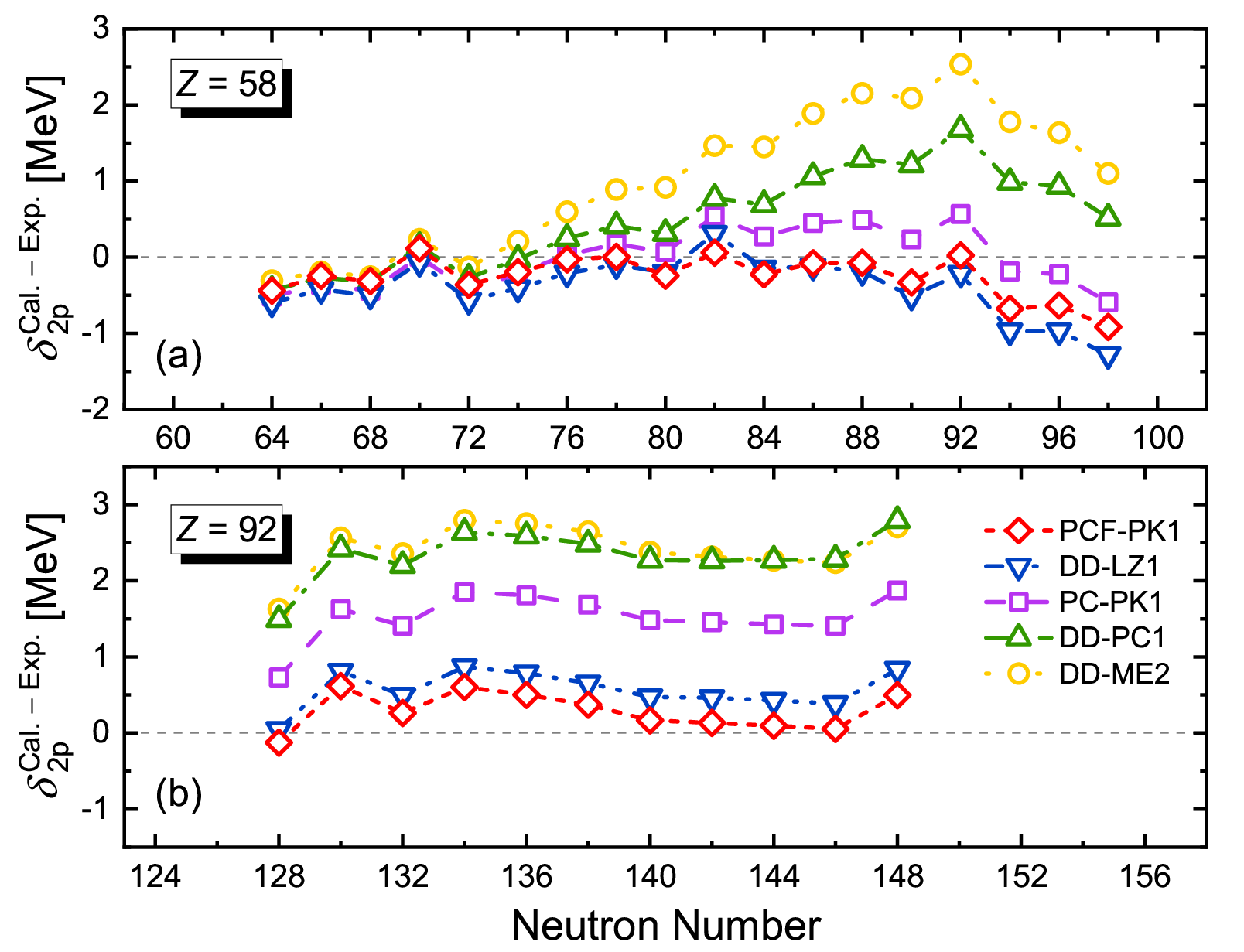}
    \caption{(a) Deviations of $\delta_{\rm 2p}$ of Ce isotopes against the experiment \cite{M.Wang_2021_CPC}, where the density functionals include PCF-PK1 (red), DD-LZ1 (blue) \cite{B.Wei_2020_CPC}, PC-PK1 (purple) \cite{P.W.Zhao_2010_PRC}, DD-PC1 (green) \cite{T.Niksic_2008_PRC}, and DD-ME2 (yellow) \cite{G.A.Lalazissis_2005_PRC}. (b) The same as (a), but for U isotopes.}
    \label{fig-8}
\end{figure}

To microscopically examine the removal of spurious shell closures, we compare the deviations of the two-proton shell gap $\delta_{\rm 2p}$ for several density functionals against the experiment \cite{M.Wang_2021_CPC}. The comparison includes PCF-PK1, DD-LZ1 \cite{B.Wei_2020_CPC}, PC-PK1 \cite{P.W.Zhao_2010_PRC}, DD-PC1 \cite{T.Niksic_2008_PRC}, and DD-ME2 \cite{G.A.Lalazissis_2005_PRC}, all calculated within the RHB framework. Fig. \ref{fig-8}(a) shows $\delta_{\rm 2p}$ deviations for Ce isotopes. The deviations for DD-ME2, DD-PC1 and PC-PK1 increase with the neutron number increasing, with peak discrepancies reaching 2.5 MeV (DD-ME2) and 1.7 MeV (DD-PC1). In contrast, the deviations for PCF-PK1 and DD-LZ1 remain small. Fig. \ref{fig-8}(b) shows $\delta_{\rm 2p}$ deviations for U isotopes. DD-ME2, DD-PC1 and PC-PK1 systematically overestimate $\delta_{\rm 2p}$, with maximum deviations reaching 2.8 MeV for DD-ME2 and DD-PC1, and 1.9 MeV for PC-PK1. While the predictions of PCF-PK1 and DD-LZ1 remain in good agreement with the experiment. The consistently small deviations of PCF-PK1 at $Z=58$ and 92 provide clear evidence that the spurious shell closures at these proton numbers have been successfully eliminated.

To further clarify the elimination of spurious shells, we examine the underlying single-particle energies. The relevant single-particle energy gaps, $E_{\rm shell}$, are defined as the energy difference between the orbitals constituting the shell closure, for example, the gap at $Z = 58$ corresponds to the energy difference between $\pi 2d_{5/2}$ and $\pi 1g_{7/2}$ levels. Table \ref{tab-2} lists the $E_{\rm shell}$ at $Z=50$ and 58 for $^{132}$Sn and $^{140}$Ce. For the doubly magic nucleus $^{132}$Sn, PCF-PK1 yields the $E_{\rm shell}$ at both $Z=50$ and $Z=58$ in agreement with experiment. However, PC-PK1, DD-PC1, and DD-ME2 overestimate $E_{\rm shell}$ at $Z=58$ by more than 1 MeV. DD-LZ1, which aims at eliminating the spurious shells in traditional relativistic EDFs \cite{B.Wei_2020_CPC},  describes a small $E_{\rm shell}$ at $Z=58$, whereas it overestimates the value at $Z=50$. For $^{140}$Ce, other functionals maintain this overestimated trend, leading to an artificial shell structure at $Z=58$. On the contrary, PCF-PK1 predicts a small $E_{\rm shell}$ at $Z=58$ for this nucleus, continuing the trend of a small $Z=58$ shell closure in its prediction for $^{132}$Sn. Table \ref{tab-2} also shows the $E_{\rm shell}$ at $Z = 82$ and 92 for $^{208}$Pb and $^{218}$U. For the doubly magic nucleus $^{208}$Pb, PCF-PK1 yields accurate results for both $E_{\rm shell}$, with deviations below 0.3 MeV. Although PC-PK1, DD-PC1 and DD-ME2 reasonably describe the $Z=82$ gap, they systematically overestimate the gap at $Z=92$, and this behavior remains for $^{218}$U. PCF-PK1 mitigates this issue by providing an improved single-particle spectrum. While DD-LZ1 gives a $\delta_{\rm 2p}$ description similar to PCF-PK1 as seen in Fig. \ref{fig-8}, its $E_{\rm shell}$ is smaller relative to other functionals but remains higher than the PCF-PK1 result as shown in Table \ref{tab-2}. The accurate description of the small $E_{\rm shell}$ at $Z=58$ and $92$ in $^{140}$Ce and $^{218}$U provides direct evidence for the successful removal of the spurious shells in the PCF-PK1 functional.

\begin{table}[htb]
    \caption{Calculated proton single-particle energy gaps $E_{\rm shell}$ at $Z = 50$ and $58$ for $^{132}$Sn and $^{140}$Ce, and at $Z = 82$ and $92$ for $^{208}$Pb and $^{218}$U. The first column under $N=82$ isotones is $E_{\rm shell}$ at $Z = 50$ and the second one is that at $Z=58$. The first column under $N=126$ isotones is $E_{\rm shell}$ at $Z = 82$ and the second one is that at $Z=92$. The values of doubly magic nuclei extracted from the experiment \cite{V.I.Isakov_2002_EPJA} are shown for comparison.}
    \begin{ruledtabular}
    \begin{tabular}{lcccccccc}
        \multirow{2}{*}{$E_{\rm shell}$ [MeV]} & \multicolumn{2}{c}{$^{132}$Sn} & \multicolumn{2}{c}{$^{140}$Ce} & \multicolumn{2}{c}{$^{208}$Pb} & \multicolumn{2}{c}{$^{218}$U} \\
        \cline{2-9}
                & 50   & 58   & 50   & 58   & 82   & 92   & 82   & 92   \\
        \hline
        PCF-PK1 & 6.25 & 0.22 & 5.38 & 0.87 & 4.06 & 1.23 & 3.93 & 1.73 \\
        DD-LZ1 & 7.29 & 1.35 & 6.71 & 2.12 & 5.20 & 1.55 & 5.01 & 2.23 \\
        PC-PK1 & 6.23 & 2.04 & 5.85 & 2.57 & 4.09 & 2.57 & 4.08 & 3.00 \\
        DD-PC1 & 6.21 & 2.27 & 5.83 & 2.65 & 3.85 & 2.96 & 4.01 & 3.27 \\
        DD-ME2 & 6.45 & 2.39 & 6.02 & 3.06 & 4.06 & 2.75 & 4.08 & 3.29 \\
        \hline 
        \bf Exp. & \bf 6.08 & \bf 0.96 & & & \bf 4.21 & \bf 0.90 & & \\ 
    \end{tabular}
    \end{ruledtabular}
    \label{tab-2}
\end{table}

The successful removal of spurious shells at $Z=58$ and 92 is directly evidenced by a more accurate description of nuclear binding energies, notably for nuclei in the affected regions such as $^{140}$Ce and $^{218}$U \cite{L.Geng_2005_PTP}. Table \ref{tab-3} compares the calculated binding energies of these nuclei with the experiment \cite{M.Wang_2021_CPC}. The table shows that for both nuclei, PCF-PK1 and DD-LZ1 yield binding energies in better agreement with the experimental data, whereas functionals PC-PK1, DD-PC1 and DD-ME2 systematically overestimate them.
\begin{table}[bt]
    \caption{Calculated binding energies of $^{140}$Ce and $^{218}$U are listed in comparison with the experiment \cite{M.Wang_2021_CPC}.}
    \begin{ruledtabular}
    % \begin{tabular}{p{2cm}p{2cm}c}
    \begin{tabular}{lrr}
        $E_{\rm b}$ [MeV] & $^{140}$Ce & $^{218}$U \\
        \hline
        PCF-PK1 & 1173.5 & 1667.5 \\
        DD-LZ1 & 1172.0 & 1666.6 \\
        PC-PK1 & 1174.4 & 1670.2 \\
        DD-PC1 & 1178.4 & 1673.5 \\
        DD-ME2 & 1175.4 & 1673.0 \\
        \hline
        \bf Exp. & \bf 1172.7 & \bf 1665.6 \\
    \end{tabular}
    \end{ruledtabular}
    \label{tab-3}
\end{table}

Ref. \cite{J.Geng_2019_PRC} shows that including the $\rho$-tensor coupling alters the density dependence of the $\sigma$-S and $\omega$-V coupling constants, which helps restore the pseudospin symmetry. Similarly, the DD-LZ1 functional achieves this restoration by directly modifying their density dependence behavior \cite{B.Wei_2020_CPC}. The novel functional PCF-PK1 incorporates both the tensor coupling and an optimized density dependence for $\sigma$-S and $\omega$-V couplings, which is key to to its success in eliminating spurious shells.

\subsection{Superheavy nuclei Region}

\begin{figure}[tb]
    \includegraphics[width=0.48\textwidth]{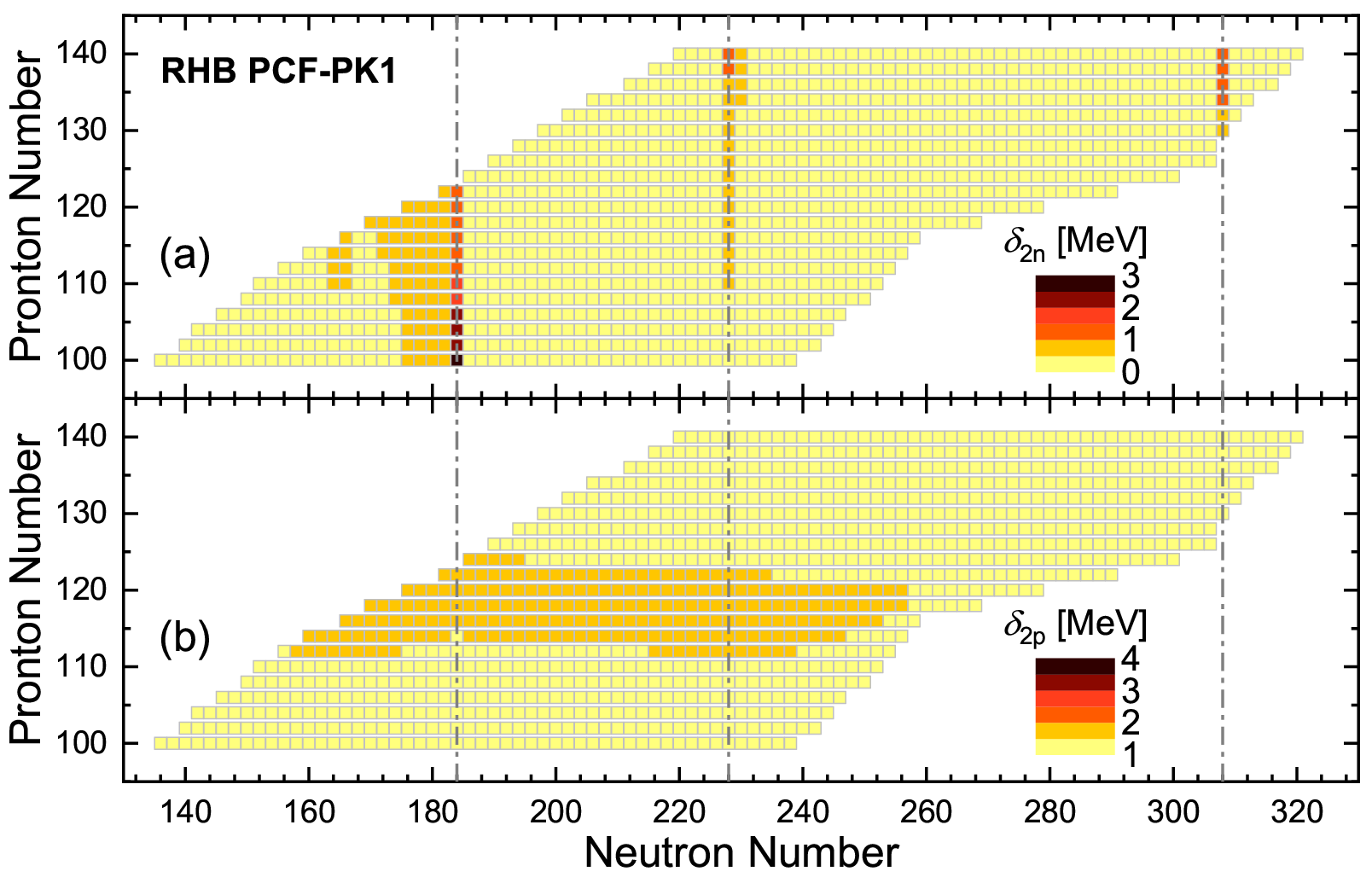}
    \caption{(a) Two-neutron and (b) two-proton shell gaps in the superheavy nuclei region with $100 \leq Z \leq 140$ calculated within the PCF-PK1 functional.}
    \label{fig-9}
\end{figure}

The search for the island of stability of superheavy nuclei (SHN) remains a long-standing challenge, where predicting the underlying magic numbers is critical \cite{G.W.Frederick_1958_PR}. Here, we employ the new functional PCF-PK1 to provide insights into this problem by examining shell gaps and single-particle spectra. Fig. \ref{fig-9}(a) presents the calculated $\delta_{\rm 2n}$ in the SHN region. PCF-PK1 predicts shell closures at $N=184$, 228, and 308. The $N=184$ gap is also robustly predicted by other functionals such as DD-PC1 \cite{T.Niksic_2008_PRC}, DD-ME2 \cite{G.A.Lalazissis_2005_PRC}, PC-PK1 \cite{P.W.Zhao_2010_PRC}, and PKA1 \cite{W.H.Long_2007_PRC}. The neutron shell closures at $N=228$ and 308 are less pronounced in our calculations. We note that the $N=228$ shell is commonly predicted in Ref. \cite{X.W.Xia_2018_ADNDT, J.J.Li_2014_PLB}, and $N=308$ is suggested by the Koura-Yamada potential model \cite{K.Hiroyuki_2013_JPSP}. However, the proton shell structure shows a notable difference. As shown in Fig. \ref{fig-9}(b), the proton shell gaps around $Z=120$ are less pronounced in PCF-PK1, with the largest value in this region being only 1.8 MeV, which is weaker than predictions from other functionals. Furthermore, the relative high proton pairing energies in this region within PCF-PK1 do not support a strong shell closure at $Z=120$, in contrast to several other models.

\begin{figure}[tb]
    \includegraphics[width=0.48\textwidth]{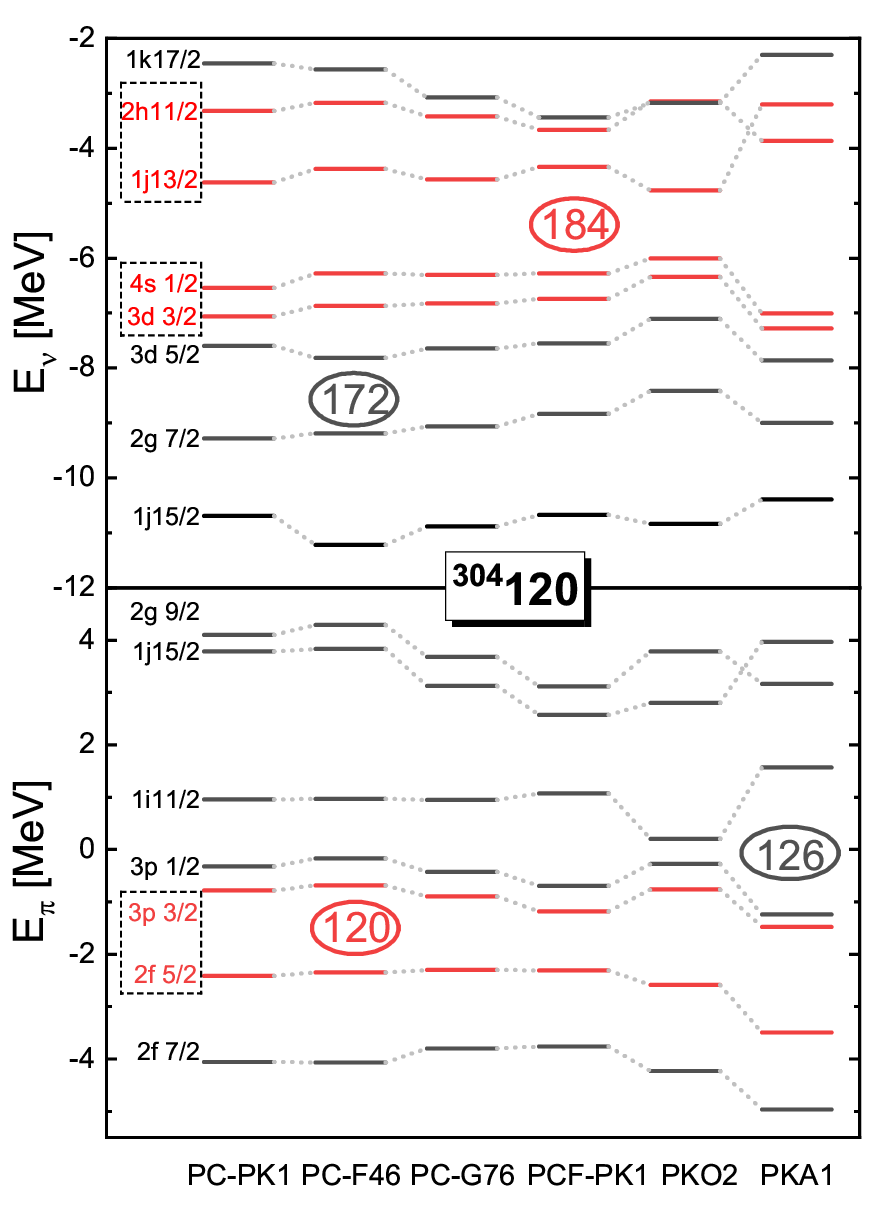}
    \caption{Single particle spectra of $^{304}$120 obtained with PCF-PK1, PC-PK1, PC-F46, PC-G76, PKO2 \cite{W.H.Long_2008_EPL}, and PKA1 \cite{W.H.Long_2007_PRC} are shown for comparison. Neutron ($E_{\nu}$) and proton ($E_{\pi}$) spectra are plotted in the upper and lower panel, respectively. The PS doublets contribute to $N=184$ and $Z=120$ are marked in red.}
    \label{fig-10}
\end{figure}

Taking $^{304}$120 as an example, Fig. \ref{fig-10} shows the single-particle energy levels calculated within the RHB framework using the PCF-PK1, PC-PK1, PC-F46, PC-G76 functionals, and within the RHFB framework using the PKO2 \cite{W.H.Long_2008_EPL}, and PKA1 \cite{W.H.Long_2007_PRC} functionals. Here, PC-F46 and PC-G76 stand for the optimal parameter fits for Dirac mass $M_{\rm D}^*=0.58 M$ and $0.70 M$, with tensor coupling strengths of 0.0458 and 1.6108, respectively \cite{Q.Zhao_2022_PRC}. This sequence illustrates a progression of increasing tensor coupling strength across the RHB calculations. For the neutron shell closure, $N=184$ is well reproduced by most functionals except PKO2. This gap is largely governed by the spin-orbit (SO) splitting of high-$j$ doublet $( 1j_{13/2}, 1j_{15/2} )$. Here, PKA1 produces the largest SO splitting, leading to the most pronounced $N=184$ gap among the functionals. Besides, $N=184$ closure is also influenced by the degeneracy of two pseudospin (PS) doublets $( 2h_{11/2}, 1j_{13/2} )$ and $( 4s_{1/2}, 3d_{3/2} )$. For the latter, the PS doublet is nearly degenerate for all functionals considered. While for the former, PCF-PK1 and PKA1 provide smaller pseudospin orbital (PSO) splittings ($\Delta E_{\rm PSO}$), in contrast to the result of PKO2. For protons, the potential shell closure at $Z = 120$ is controlled by the PS doublet $( 3p_{3/2}, 2f_{5/2} )$. The inclusion of tensor coupling in PCF-PK1 reduces $\Delta E_{\rm PSO}$ for this doublet, resulting in a less pronounced shell gap at $Z=120$. Furthermore, across the RHB calculations with PC-F46, PC-G76, and PCF-PK1, $\Delta E_{\rm PSO}$ of this doublet gradually decreases with increasing tensor coupling strength. This trend is consistent with the findings of Ref. \cite{S.W.Chen_2016_SciChina}.

\section{Summary}\label{sec-5}

In summary, we have performed systematic calculations for nuclei within the RHB theory using the functional PCF-PK1. The functional effectively incorporates exchange terms via the Fierz transformation and explicitly includes the tensor coupling. As a first step under the spherical approximation, 7210 nuclei with $8 \leq Z \leq 120$ are predicted to be bound. The rms deviation of the calculated binding energies from the 2351 measured nuclei is 7.170 MeV. For the limits of the nuclear landscape, the rms deviations of the predicted neutron and proton drip lines relative to the experiment are 2.02 and 3.46, respectively. The rms deviation of charge radii is 0.0362 fm, and it is expected to decrease with the inclusion of the deformation effect. For $\alpha$-decay energies, the achieved accuracy is 1.6263 MeV. The improved description around magic numbers indicates PCF-PK1 provides a more realistic shell structures.

Analysis of two-nucleon shell gaps confirms that the spurious shells at $Z = 58$ and $92$ are clearly eliminated with the PCF-PK1 functional. This removal consequently improves the previously overestimated binding energies of nuclei, e.g., $^{140}$Ce and $^{218}$U. This success is achieved through the combination of the inclusion of tensor coupling and an optimized density dependence for the coupling constants, and this mechanism is also employed to remove spurious shells in the framework of RHF using PKA1, and in the framework of RMF using DD-LZ1.

In the SHN region, PCF-PK1 supports neutron magic numbers at $N = 184$, 228 and 308, while the proton shell closure around $Z=120$ is less pronounced. Taking $^{304}$120 as an example, the inclusion of tensor coupling in the fitting protocols leads to a reduced $\Delta E_{\rm PSO}$. This mechanism likely explains the weaker proton shell closures predicted by PCF-PK1.

% \section*{Acknowledgments} 

% The authors thank Y. L. Yang for helpful suggestions. 

\bibliography{manuscript_temp_Notes.bib}

@article{A.V.Afanasjev_2013_PLB,
  title    = {Nuclear landscape in covariant density functional theory},
  journal  = {Physics Letters B},
  volume   = {726},
  number   = {4},
  pages    = {680-684},
  year     = {2013},
  issn     = {0370-2693},
  doi      = {https://doi.org/10.1016/j.physletb.2013.09.017},
  url      = {https://www.sciencedirect.com/science/article/pii/S037026931300734X},
  author   = {A.V. Afanasjev and S.E. Agbemava and D. Ray and P. Ring}
}

@article{B.Wei_2020_CPC,
  doi       = {10.1088/1674-1137/44/7/074107},
  url       = {https://dx.doi.org/10.1088/1674-1137/44/7/074107},
  year      = {2020},
  month     = {jul},
  publisher = {Chinese Physical Society and the Institute of High Energy Physics of the Chinese Academy of Sciences and the Institute of Modern Physics of the Chinese Academy of Sciences and IOP Publishing Ltd},
  volume    = {44},
  number    = {7},
  pages     = {074107},
  author    = {Bin Wei and Qiang Zhao and Zhi-Heng Wang and Jing Geng and Bao-Yuan Sun and Yi-Fei Niu and Wen-Hui Long},
  title     = {Novel relativistic mean field Lagrangian guided by pseudo-spin symmetry restoration *},
  journal   = {Chinese Physics C}
}

@article{D.Hirata_1997_NPA,
  title   = {A systematic study of even-even nuclei up to the drip lines within the relativistic mean field framework},
  journal = {Nuclear Physics A},
  volume  = {616},
  number  = {1},
  pages   = {438-445},
  year    = {1997},
  note    = {Radioactive Nuclear Beams},
  issn    = {0375-9474},
  doi     = {https://doi.org/10.1016/S0375-9474(97)00115-2},
  url     = {https://www.sciencedirect.com/science/article/pii/S0375947497001152},
  author  = {D. Hirata and K. Sumiyoshi and I. Tanihata and Y. Sugahara and T. Tachibana and H. Toki}
}

@article{D.S.Ahn_2019_PRL,
  title     = {Location of the Neutron Dripline at Fluorine and Neon},
  author    = {Ahn, D. S. and Fukuda, N. and Geissel, H. and Inabe, N. and Iwasa, N. and Kubo, T. and Kusaka, K. and Morrissey, D. J. and Murai, D. and Nakamura, T. and Ohtake, M. and Otsu, H. and Sato, H. and Sherrill, B. M. and Shimizu, Y. and Suzuki, H. and Takeda, H. and Tarasov, O. B. and Ueno, H. and Yanagisawa, Y. and Yoshida, K.},
  journal   = {Phys. Rev. Lett.},
  volume    = {123},
  issue     = {21},
  pages     = {212501},
  numpages  = {6},
  year      = {2019},
  month     = {Nov},
  publisher = {American Physical Society},
  doi       = {10.1103/PhysRevLett.123.212501},
  url       = {https://link.aps.org/doi/10.1103/PhysRevLett.123.212501}
}

@article{G.A.Lalazissis_1999_ADNDT,
  title   = {GROUND-STATE PROPERTIES OF EVEN-EVEN NUCLEI IN THE RELATIVISTIC MEAN-FIELD THEORY},
  journal = {Atomic Data and Nuclear Data Tables},
  volume  = {71},
  number  = {1},
  pages   = {1-40},
  year    = {1999},
  issn    = {0092-640X},
  doi     = {https://doi.org/10.1006/adnd.1998.0795},
  url     = {https://www.sciencedirect.com/science/article/pii/S0092640X98907951},
  author  = {G.A. Lalazissis and S. Raman and P. Ring}
}

@article{G.A.Lalazissis_2005_PRC,
  title     = {New relativistic mean-field interaction with density-dependent meson-nucleon couplings},
  author    = {Lalazissis, G. A. and Nik\ifmmode \check{s}\else \v{s}\fi{}i\ifmmode \acute{c}\else \'{c}\fi{}, T. and Vretenar, D. and Ring, P.},
  journal   = {Phys. Rev. C},
  volume    = {71},
  issue     = {2},
  pages     = {024312},
  numpages  = {10},
  year      = {2005},
  month     = {Feb},
  publisher = {American Physical Society},
  doi       = {10.1103/PhysRevC.71.024312},
  url       = {https://link.aps.org/doi/10.1103/PhysRevC.71.024312}
}

@article{I.Angeli_2013_ADNDT,
  title    = {Table of experimental nuclear ground state charge radii: An update},
  journal  = {Atomic Data and Nuclear Data Tables},
  volume   = {99},
  number   = {1},
  pages    = {69-95},
  year     = {2013},
  issn     = {0092-640X},
  doi      = {https://doi.org/10.1016/j.adt.2011.12.006},
  url      = {https://www.sciencedirect.com/science/article/pii/S0092640X12000265},
  author   = {I. Angeli and K.P. Marinova}
}

@article{J.Erler_2012_Nature,
  author  = {Erler, Jochen
             and Birge, Noah
             and Kortelainen, Markus
             and Nazarewicz, Witold
             and Olsen, Erik
             and Perhac, Alexander M.
             and Stoitsov, Mario},
  title   = {The limits of the nuclear landscape},
  journal = {Nature},
  year    = {2012},
  month   = {Jun},
  day     = {01},
  volume  = {486},
  number  = {7404},
  pages   = {509-512},
  issn    = {1476-4687},
  doi     = {10.1038/nature11188},
  url     = {https://doi.org/10.1038/nature11188}
}

@article{J.Geng_2019_PRC,
  title     = {Pseudospin symmetry restoration and the in-medium balance between nuclear attractive and repulsive interactions},
  author    = {Geng, Jing and Li, Jia Jie and Long, Wen Hui and Niu, Yi Fei and Chang, Shi Yao},
  journal   = {Phys. Rev. C},
  volume    = {100},
  issue     = {5},
  pages     = {051301},
  numpages  = {6},
  year      = {2019},
  month     = {Nov},
  publisher = {American Physical Society},
  doi       = {10.1103/PhysRevC.100.051301},
  url       = {https://link.aps.org/doi/10.1103/PhysRevC.100.051301}
}

@article{K.Blaum_2006_PR,
  title    = {High-accuracy mass spectrometry with stored ions},
  journal  = {Physics Reports},
  volume   = {425},
  number   = {1},
  pages    = {1-78},
  year     = {2006},
  issn     = {0370-1573},
  doi      = {https://doi.org/10.1016/j.physrep.2005.10.011},
  url      = {https://www.sciencedirect.com/science/article/pii/S0370157305004643},
  author   = {Klaus Blaum}
}

@article{K.Y.Zhang_2022_ADNDT,
  title   = {Nuclear mass table in deformed relativistic Hartree-Bogoliubov theory in continuum, I: Even-even nuclei},
  journal = {Atomic Data and Nuclear Data Tables},
  volume  = {144},
  pages   = {101488},
  year    = {2022},
  issn    = {0092-640X},
  doi     = {https://doi.org/10.1016/j.adt.2022.101488},
  url     = {https://www.sciencedirect.com/science/article/pii/S0092640X22000018},
  author  = {Kaiyuan Zhang and Myung-Ki Cheoun and Yong-Beom Choi and Pooi Seong Chong and Jianmin Dong and Zihao Dong and Xiaokai Du and Lisheng Geng and Eunja Ha and Xiao-Tao He and Chan Heo and Meng Chit Ho and Eun Jin In and Seonghyun Kim and Youngman Kim and Chang-Hwan Lee and Jenny Lee and Hexuan Li and Zhipan Li and Tianpeng Luo and Jie Meng and Myeong-Hwan Mun and Zhongming Niu and Cong Pan and Panagiota Papakonstantinou and Xinle Shang and Caiwan Shen and Guofang Shen and Wei Sun and Xiang-Xiang Sun and Chi Kin Tam and  Thaivayongnou and Chen Wang and Xingzhi Wang and Sau Hei Wong and Jiawei Wu and Xinhui Wu and Xuewei Xia and Yijun Yan and Ryan Wai-Yen Yeung and To Chung Yiu and Shuangquan Zhang and Wei Zhang and Xiaoyan Zhang and Qiang Zhao and Shan-Gui Zhou}
}

@article{L.Geng_2005_PTP,
  author  = {Geng, Lisheng and Toki, Hiroshi and Meng, Jie},
  title   = {{Masses, Deformations and Charge Radii—Nuclear Ground-State Properties in the Relativistic Mean Field Model}},
  journal = {Progress of Theoretical Physics},
  volume  = {113},
  number  = {4},
  pages   = {785-800},
  year    = {2005},
  month   = {04},
  issn    = {0033-068X},
  doi     = {10.1143/PTP.113.785},
  url     = {https://doi.org/10.1143/PTP.113.785},
}

@article{M.Thoennessen_2004_RPP,
  doi       = {10.1088/0034-4885/67/7/R04},
  url       = {https://dx.doi.org/10.1088/0034-4885/67/7/R04},
  year      = {2004},
  month     = {jun},
  publisher = {},
  volume    = {67},
  number    = {7},
  pages     = {1187},
  author    = {M Thoennessen},
  title     = {Reaching the limits of nuclear stability},
  journal   = {Reports on Progress in Physics}
}

@article{M.Wang_2012_CPC,
  doi       = {10.1088/1674-1137/36/12/003},
  url       = {https://dx.doi.org/10.1088/1674-1137/36/12/003},
  year      = {2012},
  month     = {dec},
  publisher = {},
  volume    = {36},
  number    = {12},
  pages     = {1603},
  author    = {M .Wang and G. Audi and A. H. Wapstra and F. G. Kondev and M. MacCormick and X. Xu and B. Pfeiffer},
  title     = {The Ame2012 atomic mass evaluation},
  journal   = {Chinese Physics C}
}

@article{M.Wang_2021_CPC,
  doi       = {10.1088/1674-1137/abddaf},
  url       = {https://dx.doi.org/10.1088/1674-1137/abddaf},
  year      = {2021},
  month     = {mar},
  publisher = {Chinese Physical Society and the Institute of High Energy Physics of the Chinese Academy of Sciences and the Institute of Modern Physics of the Chinese Academy of Sciences and IOP Publishing Ltd},
  volume    = {45},
  number    = {3},
  pages     = {030003},
  author    = {Meng Wang and W.J. Huang and F.G. Kondev and G. Audi and S. Naimi},
  title     = {The AME 2020 atomic mass evaluation (II). Tables, graphs and references*},
  journal   = {Chinese Physics C}
}

@article{N.Wang_2014_PLB,
  title   = {Surface diffuseness correction in global mass formula},
  journal = {Physics Letters B},
  volume  = {734},
  pages   = {215-219},
  year    = {2014},
  issn    = {0370-2693},
  doi     = {https://doi.org/10.1016/j.physletb.2014.05.049},
  url     = {https://www.sciencedirect.com/science/article/pii/S037026931400358X},
  author  = {Ning Wang and Min Liu and Xizhen Wu and Jie Meng}
}

@article{P.Moller_2016_ADNDT,
    title = {Nuclear ground-state masses and deformations: FRDM(2012)},
    journal = {Atomic Data and Nuclear Data Tables},
    volume = {109-110},
    pages = {1-204},
    year = {2016},
    issn = {0092-640X},
    doi = {https://doi.org/10.1016/j.adt.2015.10.002},
    url = {https://www.sciencedirect.com/science/article/pii/S0092640X1600005X},
    author = {P. M\"oller and A.J. Sierk and T. Ichikawa and H. Sagawa},
}

@book{P.Ring_many-body_1980,
  address   = {New York},
  author    = {Ring, P. and Schuck, P.},
  publisher = {Springer-Verlag},
  title     = {The nuclear many-body problem},
  year      = {1980}
}

@article{P.W.Zhao_2010_PRC,
  title     = {New parametrization for the nuclear covariant energy density functional with a point-coupling interaction},
  author    = {Zhao, P. W. and Li, Z. P. and Yao, J. M. and Meng, J.},
  journal   = {Phys. Rev. C},
  volume    = {82},
  issue     = {5},
  pages     = {054319},
  numpages  = {14},
  year      = {2010},
  month     = {Nov},
  publisher = {American Physical Society},
  doi       = {10.1103/PhysRevC.82.054319},
  url       = {https://link.aps.org/doi/10.1103/PhysRevC.82.054319}
}

@article{Q.Zhao_2022_PRC,
  title     = {Covariant density functional theory with localized exchange terms},
  author    = {Zhao, Qiang and Ren, Zhengxue and Zhao, Pengwei and Meng, Jie},
  journal   = {Phys. Rev. C},
  volume    = {106},
  issue     = {3},
  pages     = {034315},
  numpages  = {15},
  year      = {2022},
  month     = {Sep},
  publisher = {American Physical Society},
  doi       = {10.1103/PhysRevC.106.034315},
  url       = {https://link.aps.org/doi/10.1103/PhysRevC.106.034315}
}

@article{S.E.Agbemava_2014_PRC,
  title     = {Global performance of covariant energy density functionals: Ground state observables of even-even nuclei and the estimate of theoretical uncertainties},
  author    = {Agbemava, S. E. and Afanasjev, A. V. and Ray, D. and Ring, P.},
  journal   = {Phys. Rev. C},
  volume    = {89},
  issue     = {5},
  pages     = {054320},
  numpages  = {37},
  year      = {2014},
  month     = {May},
  publisher = {American Physical Society},
  doi       = {10.1103/PhysRevC.89.054320},
  url       = {https://link.aps.org/doi/10.1103/PhysRevC.89.054320}
}

@article{S.Goriely_2013_PRC,
  title     = {Hartree-Fock-Bogoliubov nuclear mass model with 0.50 MeV accuracy based on standard forms of Skyrme and pairing functionals},
  author    = {Goriely, S. and Chamel, N. and Pearson, J. M.},
  journal   = {Phys. Rev. C},
  volume    = {88},
  issue     = {6},
  pages     = {061302},
  numpages  = {5},
  year      = {2013},
  month     = {Dec},
  publisher = {American Physical Society},
  doi       = {10.1103/PhysRevC.88.061302},
  url       = {https://link.aps.org/doi/10.1103/PhysRevC.88.061302}
}

@article{S.W.Chen_2016_SciChina,
  author  = {Chen, ShouWan
             and Li, DongPeng
             and Guo, JianYou},
  title   = {Tensor coupling effect on relativistic symmetries},
  journal = {Science China Physics, Mechanics \& Astronomy},
  year    = {2016},
  month   = {May},
  day     = {25},
  volume  = {59},
  number  = {8},
  pages   = {682011},
  issn    = {1869-1927},
  doi     = {10.1007/s11433-016-0042-5},
  url     = {https://doi.org/10.1007/s11433-016-0042-5}
}

@article{T.Li_2021_ADNDT,
  title    = {Compilation of recent nuclear ground state charge radius measurements and tests for models},
  journal  = {Atomic Data and Nuclear Data Tables},
  volume   = {140},
  pages    = {101440},
  year     = {2021},
  issn     = {0092-640X},
  doi      = {https://doi.org/10.1016/j.adt.2021.101440},
  url      = {https://www.sciencedirect.com/science/article/pii/S0092640X21000267},
  author   = {Tao Li and Yani Luo and Ning Wang}
}

@article{T.Niksic_2008_PRC,
  title     = {Relativistic nuclear energy density functionals: Adjusting parameters to binding energies},
  author    = {Nik\ifmmode \check{s}\else \v{s}\fi{}i\ifmmode \acute{c}\else \'{c}\fi{}, T. and Vretenar, D. and Ring, P.},
  journal   = {Phys. Rev. C},
  volume    = {78},
  issue     = {3},
  pages     = {034318},
  numpages  = {19},
  year      = {2008},
  month     = {Sep},
  publisher = {American Physical Society},
  doi       = {10.1103/PhysRevC.78.034318},
  url       = {https://link.aps.org/doi/10.1103/PhysRevC.78.034318}
}

@article{T.Niksic_2014_CPC,
  title   = {DIRHB—A relativistic self-consistent mean-field framework for atomic nuclei},
  journal = {Computer Physics Communications},
  volume  = {185},
  number  = {6},
  pages   = {1808-1821},
  year    = {2014},
  issn    = {0010-4655},
  doi     = {https://doi.org/10.1016/j.cpc.2014.02.027},
  url     = {https://www.sciencedirect.com/science/article/pii/S0010465514000836},
  author  = {T. Nik\v{s}i\'{c} and N. Paar and D. Vretenar and P. Ring}
}

@article{V.I.Isakov_2002_EPJA,
  author  = {V.I. Isakov and K.I. Erokhina and H. Mach and M. Sanchez-Vega and B. Fogelberg},
  title   = {On the difference between proton and neutron spin-orbit splittings in nuclei},
  doi     = {10.1140/epja/iepja1393},
  url     = {https://doi.org/10.1140/epja/iepja1393},
  journal = {Eur. Phys. J. A},
  year    = {2002},
  volume  = {14},
  number  = {1},
  pages   = {29-36},
  month   = {}
}

@article{W.H.Long_2007_PRC,
  title     = {Shell structure and \ensuremath{\rho}-tensor correlations in density dependent relativistic Hartree-Fock theory},
  author    = {Long, WenHui and Sagawa, Hiroyuki and Giai, Nguyen Van and Meng, Jie},
  journal   = {Phys. Rev. C},
  volume    = {76},
  issue     = {3},
  pages     = {034314},
  numpages  = {11},
  year      = {2007},
  month     = {Sep},
  publisher = {American Physical Society},
  doi       = {10.1103/PhysRevC.76.034314},
  url       = {https://link.aps.org/doi/10.1103/PhysRevC.76.034314}
}

@article{W.H.Long_2008_EPL,
  doi       = {10.1209/0295-5075/82/12001},
  url       = {https://dx.doi.org/10.1209/0295-5075/82/12001},
  year      = {2008},
  month     = {mar},
  publisher = {},
  volume    = {82},
  number    = {1},
  pages     = {12001},
  author    = {WenHui Long and Hiroyuki Sagawa and Jie Meng and Nguyen Van Giai},
  title     = {Evolution of nuclear shell structure due to the pion exchange potential},
  journal   = {Europhysics Letters}
}

@article{X.W.Xia_2018_ADNDT,
  title    = {The limits of the nuclear landscape explored by the relativistic continuum Hartree-Bogoliubov theory},
  journal  = {Atomic Data and Nuclear Data Tables},
  volume   = {121-122},
  pages    = {1-215},
  year     = {2018},
  issn     = {0092-640X},
  doi      = {https://doi.org/10.1016/j.adt.2017.09.001},
  url      = {https://www.sciencedirect.com/science/article/pii/S0092640X17300451},
  author   = {X.W. Xia and Y. Lim and P.W. Zhao and H.Z. Liang and X.Y. Qu and Y. Chen and H. Liu and L.F. Zhang and S.Q. Zhang and Y. Kim and J. Meng}
}

@article{Y.L.Yang_2021_PRC,
  title     = {Nuclear landscape in a mapped collective Hamiltonian from covariant density functional theory},
  author    = {Yang, Y. L. and Wang, Y. K. and Zhao, P. W. and Li, Z. P.},
  journal   = {Phys. Rev. C},
  volume    = {104},
  issue     = {5},
  pages     = {054312},
  numpages  = {6},
  year      = {2021},
  month     = {Nov},
  publisher = {American Physical Society},
  doi       = {10.1103/PhysRevC.104.054312},
  url       = {https://link.aps.org/doi/10.1103/PhysRevC.104.054312}
}

@article{Y.Tian_2009_PLB,
  title   = {A finite range pairing force for density functional theory in superfluid nuclei},
  journal = {Physics Letters B},
  volume  = {676},
  number  = {1},
  pages   = {44-50},
  year    = {2009},
  issn    = {0370-2693},
  doi     = {https://doi.org/10.1016/j.physletb.2009.04.067},
  url     = {https://www.sciencedirect.com/science/article/pii/S0370269309004912},
  author  = {Y. Tian and Z.Y. Ma and P. Ring}
}

@article{Z.Y.Zhang_2019_PRL,
  title     = {New Isotope $^{220}\mathrm{Np}$: Probing the Robustness of the $N=126$ Shell Closure in Neptunium},
  author    = {Zhang, Z. Y. and Gan, Z. G. and Yang, H. B. and Ma, L. and Huang, M. H. and Yang, C. L. and Zhang, M. M. and Tian, Y. L. and Wang, Y. S. and Sun, M. D. and Lu, H. Y. and Zhang, W. Q. and Zhou, H. B. and Wang, X. and Wu, C. G. and Duan, L. M. and Huang, W. X. and Liu, Z. and Ren, Z. Z. and Zhou, S. G. and Zhou, X. H. and Xu, H. S. and Tsyganov, Yu. S. and Voinov, A. A. and Polyakov, A. N.},
  journal   = {Phys. Rev. Lett.},
  volume    = {122},
  issue     = {19},
  pages     = {192503},
  numpages  = {6},
  year      = {2019},
  month     = {May},
  publisher = {American Physical Society},
  doi       = {10.1103/PhysRevLett.122.192503},
  url       = {https://link.aps.org/doi/10.1103/PhysRevLett.122.192503}
}

@article{P.Guo_2024_ADNDT,
    title = {Nuclear mass table in deformed relativistic Hartree–Bogoliubov theory in continuum, II: Even-Z nuclei},
    journal = {Atomic Data and Nuclear Data Tables},
    volume = {158},
    pages = {101661},
    year = {2024},
    issn = {0092-640X},
    doi = {https://doi.org/10.1016/j.adt.2024.101661},
    url = {https://www.sciencedirect.com/science/article/pii/S0092640X24000263},
    author = {Peng Guo and Xiaojie Cao and Kangmin Chen and Zhihui Chen and Myung-Ki Cheoun and Yong-Beom Choi and Pak Chung Lam and Wenmin Deng and Jianmin Dong and Pengxiang Du and Xiaokai Du and Kangda Duan and Xiaohua Fan and Wei Gao and Lisheng Geng and Eunja Ha and Xiao-Tao He and Jinniu Hu and Jingke Huang and Kun Huang and Yanan Huang and Zidan Huang and Kim Da Hyung and Hoi Yat Chan and Xiaofei Jiang and Seonghyun Kim and Youngman Kim and Chang-Hwan Lee and Jenny Lee and Jian Li and Minglong Li and Zhipan Li and Zhengzheng Li and Zhanjiang Lian and Haozhao Liang and Lang Liu and Xiao Lu and Zhi-Rui Liu and Jie Meng and Ziyan Meng and Myeong-Hwan Mun and Yifei Niu and Zhongming Niu and Cong Pan and Jing Peng and Xiaoying Qu and Panagiota Papakonstantinou and Tianshuai Shang and Xinle Shang and Caiwan Shen and Guofang Shen and Tingting Sun and Xiang-Xiang Sun and Sibo Wang and Tianyu Wang and Yiran Wang and Yuanyuan Wang and Jiawei Wu and Liang Wu and Xinhui Wu and Xuewei Xia and Huihui Xie and Jiangming Yao and Kwan Yau Ip and To Chung Yiu and Jianghan Yu and Yangyang Yu and Kaiyuan Zhang and Shijie Zhang and Shuangquan Zhang and Wei Zhang and Xiaoyan Zhang and Yanxin Zhang and Ying Zhang and Yingxun Zhang and Zhenhua Zhang and Qiang Zhao and Yingchun Zhao and Ruyou Zheng and Chang Zhou and Shan-Gui Zhou and Lianjian Zou}
}

@article{Y.Yu_2024_PRL,
  title = {Nuclear Structure of Dripline Nuclei Elucidated through Precision Mass Measurements of $^{23}\mathrm{Si}$, $^{26}\mathrm{P}$, $^{27,28}\mathrm{S}$, and $^{31}\mathrm{Ar}$},
  author = {Yu, Y. and Xing, Y. M. and Zhang, Y. H. and Wang, M. and Zhou, X. H. and Li, J. G. and Li, H. H. and Yuan, Q. and Niu, Y. F. and Huang, Y. N. and Geng, J. and Guo, J. Y. and Chen, J. W. and Pei, J. C. and Xu, F. R. and Litvinov, Yu. A. and Blaum, K. and de Angelis, G. and Tanihata, I. and Yamaguchi, T. and Zhou, X. and Xu, H. S. and Chen, Z. Y. and Chen, R. J. and Deng, H. Y. and Fu, C. Y. and Ge, W. W. and Huang, W. J. and Jiao, H. Y. and Luo, Y. F. and Li, H. F. and Liao, T. and Shi, J. Y. and Si, M. and Sun, M. Z. and Shuai, P. and Tu, X. L. and Wang, Q. and Xu, X. and Yan, X. L. and Yuan, Y. J. and Zhang, M.},
  journal = {Phys. Rev. Lett.},
  volume = {133},
  issue = {22},
  pages = {222501},
  numpages = {7},
  year = {2024},
  month = {Nov},
  publisher = {American Physical Society},
  doi = {10.1103/PhysRevLett.133.222501},
  url = {https://link.aps.org/doi/10.1103/PhysRevLett.133.222501}
}

@article{Y.M.Xing_2025_PRL,
  title = {$Z=14$ Magicity Revealed by the Mass of the Proton Dripline Nucleus $^{22}\mathrm{Si}$},
  author = {Xing, Y. M. and Luo, Y. F. and Zhang, Y. H. and Wang, M. and Zhou, X. H. and Li, J. G. and Li, K. H. and Yuan, Q. and Niu, Y. F. and Guo, J. Y. and Pei, J. C. and Xu, F. R. and de Angelis, G. and Litvinov, Yu. A. and Blaum, K. and Tanihata, I. and Yamaguchi, T. and Yu, Y. and Zhou, X. and Xu, H. S. and Chen, Z. Y. and Chen, R. J. and Deng, H. Y. and Fu, C. Y. and Ge, W. W. and Huang, W. J. and Jiao, H. Y. and Li, H. F. and Liao, T. and Shi, J. Y. and Si, M. and Sun, M. Z. and Shuai, P. and Tu, X. L. and Wang, Q. and Xu, X. and Yan, X. L. and Yuan, Y. J. and Zhang, M.},
  journal = {Phys. Rev. Lett.},
  volume = {135},
  issue = {1},
  pages = {012501},
  numpages = {8},
  year = {2025},
  month = {Jul},
  publisher = {American Physical Society},
  doi = {10.1103/ffwt-n7yc},
  url = {https://link.aps.org/doi/10.1103/ffwt-n7yc}
}

@book{J.Meng_2016_RDFT,
    author = {Meng, Jie},
    title = {Relativistic Density Functional for Nuclear Structure},
    publisher = {WORLD SCIENTIFIC},
    year = {2016},
    doi = {10.1142/9872}
}

@misc{M.Bradley_nucnet,
  howpublished = {\url{https://sourceforge.net/projects/nucnet-tools/}},
}

@Article{F.Wienholtz_2013_Nature,
    author={Wienholtz, F.
    and Beck, D.
    and Blaum, K.
    and Borgmann, Ch.
    and Breitenfeldt, M.
    and Cakirli, R. B.
    and George, S.
    and Herfurth, F.
    and Holt, J. D.
    and Kowalska, M.
    and Kreim, S.
    and Lunney, D.
    and Manea, V.
    and Men{\'e}ndez, J.
    and Neidherr, D.
    and Rosenbusch, M.
    and Schweikhard, L.
    and Schwenk, A.
    and Simonis, J.
    and Stanja, J.
    and Wolf, R. N.
    and Zuber, K.},
    title={Masses of exotic calcium isotopes pin down nuclear forces},
    journal={Nature},
    year={2013},
    month={Jun},
    day={01},
    volume={498},
    number={7454},
    pages={346-349},
    issn={1476-4687},
    doi={10.1038/nature12226},
    url={https://doi.org/10.1038/nature12226}
}

@Article{B.H.Sun_2015_FP,
    author={Sun, B. H.
    and Litvinov, Yu. A.
    and Tanihata, I.
    and Zhang, Y. H.},
    title={Toward precision mass measurements of neutron-rich nuclei relevant to r-process nucleosynthesis},
    journal={Frontiers of Physics},
    year={2015},
    month={Aug},
    day={01},
    volume={10},
    number={4},
    pages={1-25},
    issn={2095-0470},
    doi={10.1007/s11467-015-0503-z},
    url={https://doi.org/10.1007/s11467-015-0503-z}
}

@article{Y.Aboussir_1995_ADNDT,
    title = {Nuclear mass formula via an approximation to the Hartree—Fock method},
    journal = {Atomic Data and Nuclear Data Tables},
    volume = {61},
    number = {1},
    pages = {127-176},
    year = {1995},
    issn = {0092-640X},
    doi = {https://doi.org/10.1016/S0092-640X(95)90014-4},
    url = {https://www.sciencedirect.com/science/article/pii/S0092640X95900144},
    author = {Y. Aboussir and J.M. Pearson and A.K. Dutta and F. Tondeur}
}

@article{H.F.Zhang_2014_NPA,
    title = {Development of metakaolin–fly ash based geopolymers for fire resistance applications},
    journal = {Construction and Building Materials},
    volume = {55},
    pages = {38-45},
    year = {2014},
    issn = {0950-0618},
    doi = {https://doi.org/10.1016/j.conbuildmat.2014.01.040},
    url = {https://www.sciencedirect.com/science/article/pii/S0950061814000646},
    author = {Hai Yan Zhang and Venkatesh Kodur and Shu Liang Qi and Liang Cao and Bo Wu}
}

@article{S.Goriely_2009_PRL_HFB17,
  title = {Skyrme-Hartree-Fock-Bogoliubov Nuclear Mass Formulas: Crossing the 0.6 MeV Accuracy Threshold with Microscopically Deduced Pairing},
  author = {Goriely, S. and Chamel, N. and Pearson, J. M.},
  journal = {Phys. Rev. Lett.},
  volume = {102},
  issue = {15},
  pages = {152503},
  numpages = {4},
  year = {2009},
  month = {Apr},
  publisher = {American Physical Society},
  doi = {10.1103/PhysRevLett.102.152503},
  url = {https://link.aps.org/doi/10.1103/PhysRevLett.102.152503}
}

@article{S.Goriely_2009_PRL_D1M,
  title = {First Gogny-Hartree-Fock-Bogoliubov Nuclear Mass Model},
  author = {Goriely, S. and Hilaire, S. and Girod, M. and P\'eru, S.},
  journal = {Phys. Rev. Lett.},
  volume = {102},
  issue = {24},
  pages = {242501},
  numpages = {4},
  year = {2009},
  month = {Jun},
  publisher = {American Physical Society},
  doi = {10.1103/PhysRevLett.102.242501},
  url = {https://link.aps.org/doi/10.1103/PhysRevLett.102.242501}
}

@article{Z.X.Liu_2023_PLB,
    title = {The optimized point-coupling interaction for the relativistic energy density functional of Hartree-Bogoliubov approach quantifying the nuclear bulk properties},
    journal = {Physics Letters B},
    volume = {842},
    pages = {137946},
    year = {2023},
    issn = {0370-2693},
    doi = {https://doi.org/10.1016/j.physletb.2023.137946},
    url = {https://www.sciencedirect.com/science/article/pii/S0370269323002800},
    author = {Zi Xin Liu and Yi Hua Lam and Ning Lu and Peter Ring}
}

@article{J.Meng_1996_PRL,
  title = {Relativistic Hartree-Bogoliubov Description of the Neutron Halo in ${}^{11}$Li},
  author = {Meng, J. and Ring, P.},
  journal = {Phys. Rev. Lett.},
  volume = {77},
  issue = {19},
  pages = {3963--3966},
  numpages = {0},
  year = {1996},
  month = {Nov},
  publisher = {American Physical Society},
  doi = {10.1103/PhysRevLett.77.3963},
  url = {https://link.aps.org/doi/10.1103/PhysRevLett.77.3963}
}

@article{J.Meng_1998_NPA,
    title = {Relativistic continuum Hartree-Bogoliubov theory with both zero range and finite range Gogny force and their application},
    journal = {Nuclear Physics A},
    volume = {635},
    number = {1},
    pages = {3-42},
    year = {1998},
    issn = {0375-9474},
    doi = {https://doi.org/10.1016/S0375-9474(98)00178-X},
    url = {https://www.sciencedirect.com/science/article/pii/S037594749800178X},
    author = {Jie Meng}
}

@article{M.Thoennessen_2004_RoPP,
    doi = {10.1088/0034-4885/67/7/R04},
    url = {https://doi.org/10.1088/0034-4885/67/7/R04},
    year = {2004},
    month = {jun},
    publisher = {},
    volume = {67},
    number = {7},
    pages = {1187},
    author = {M Thoennessen},
    title = {Reaching the limits of nuclear stability},
    journal = {Reports on Progress in Physics}
}

@article{H.Z.Liang_2012_PRC1,
  title = {Localized form of Fock terms in nuclear covariant density functional theory},
  author = {Liang, Haozhao and Zhao, Pengwei and Ring, Peter and Roca-Maza, Xavier and Meng, Jie},
  journal = {Phys. Rev. C},
  volume = {86},
  issue = {2},
  pages = {021302},
  numpages = {5},
  year = {2012},
  month = {Aug},
  publisher = {American Physical Society},
  doi = {10.1103/PhysRevC.86.021302},
  url = {https://link.aps.org/doi/10.1103/PhysRevC.86.021302}
}

@book{W.Greiner_2009_Fierz,
  author       = {Greiner, Walter and
                  Mueller, Berndt},
  title        = {Gauge theory of weak interactions. 4. ed.},
  publisher    = {Springer},
  year         = 2025,
  month        = jan,
  doi          = {10.1007/978-3-540-87843-8}
}

@article{A.Sulaksono_2003_AP,
    title = {Mapping exchange in relativistic Hartree–Fock},
    journal = {Annals of Physics},
    volume = {306},
    number = {1},
    pages = {36-57},
    year = {2003},
    issn = {0003-4916},
    doi = {https://doi.org/10.1016/S0003-4916(03)00073-3},
    url = {https://www.sciencedirect.com/science/article/pii/S0003491603000733},
    author = {A. Sulaksono and T. B\"urvenich and J.A. Maruhn and P.-G. Reinhard and W. Greiner}
}

@article{J.J.Li_2014_PLB,
    title = {Superheavy magic structures in the relativistic Hartree–Fock–Bogoliubov approach},
    journal = {Physics Letters B},
    volume = {732},
    pages = {169-173},
    year = {2014},
    issn = {0370-2693},
    doi = {https://doi.org/10.1016/j.physletb.2014.03.031},
    url = {https://www.sciencedirect.com/science/article/pii/S0370269314001877},
    author = {Jia Jie Li and Wen Hui Long and Jérôme Margueron and Nguyen {Van Giai}}
}

@article{H.Z.Liang_2015_PR,
    title = {Hidden pseudospin and spin symmetries and their origins in atomic nuclei},
    journal = {Physics Reports},
    volume = {570},
    pages = {1-84},
    year = {2015},
    note = {Hidden pseudospin and spin symmetries and their origins in atomic nuclei},
    issn = {0370-1573},
    doi = {https://doi.org/10.1016/j.physrep.2014.12.005},
    url = {https://www.sciencedirect.com/science/article/pii/S0370157315000502},
    author = {Haozhao Liang and Jie Meng and Shan-Gui Zhou}
}

@article{W.Koepf_1989_NPA,
    title = {A relativistic description of rotating nuclei: The yrast line of 20Ne},
    journal = {Nuclear Physics A},
    volume = {493},
    number = {1},
    pages = {61-82},
    year = {1989},
    issn = {0375-9474},
    doi = {https://doi.org/10.1016/0375-9474(89)90532-0},
    url = {https://www.sciencedirect.com/science/article/pii/0375947489905320},
    author = {W. Koepf and P. Ring}
}

@article{M.G.Mayer_1949_PR,
  title = {On Closed Shells in Nuclei. II},
  author = {Mayer, Maria Goeppert},
  journal = {Phys. Rev.},
  volume = {75},
  issue = {12},
  pages = {1969--1970},
  numpages = {0},
  year = {1949},
  month = {Jun},
  publisher = {American Physical Society},
  doi = {10.1103/PhysRev.75.1969},
  url = {https://link.aps.org/doi/10.1103/PhysRev.75.1969}
}

@article{O.Haxel_1949_PR,
  title = {On the "Magic Numbers" in Nuclear Structure},
  author = {Haxel, Otto and Jensen, J. Hans D. and Suess, Hans E.},
  journal = {Phys. Rev.},
  volume = {75},
  issue = {11},
  pages = {1766--1766},
  numpages = {0},
  year = {1949},
  month = {Jun},
  publisher = {American Physical Society},
  doi = {10.1103/PhysRev.75.1766.2},
  url = {https://link.aps.org/doi/10.1103/PhysRev.75.1766.2}
}

@article{A.Bohr_1958_PR,
  title = {Possible Analogy between the Excitation Spectra of Nuclei and Those of the Superconducting Metallic State},
  author = {Bohr, A. and Mottelson, B. R. and Pines, D.},
  journal = {Phys. Rev.},
  volume = {110},
  issue = {4},
  pages = {936--938},
  numpages = {0},
  year = {1958},
  month = {May},
  publisher = {American Physical Society},
  doi = {10.1103/PhysRev.110.936},
  url = {https://link.aps.org/doi/10.1103/PhysRev.110.936}
}

@article{H.Z.Liang_2008_PRL,
  title = {Spin-Isospin Resonances: A Self-Consistent Covariant Description},
  author = {Liang, Haozhao and Van Giai, Nguyen and Meng, Jie},
  journal = {Phys. Rev. Lett.},
  volume = {101},
  issue = {12},
  pages = {122502},
  numpages = {4},
  year = {2008},
  month = {Sep},
  publisher = {American Physical Society},
  doi = {10.1103/PhysRevLett.101.122502},
  url = {https://link.aps.org/doi/10.1103/PhysRevLett.101.122502}
}

@article{H.Z.Liang_2012_PRC2,
  title = {Fine structure of charge-exchange spin-dipole excitations in ${}^{16}$O},
  author = {Liang, Haozhao and Zhao, Pengwei and Meng, Jie},
  journal = {Phys. Rev. C},
  volume = {85},
  issue = {6},
  pages = {064302},
  numpages = {5},
  year = {2012},
  month = {Jun},
  publisher = {American Physical Society},
  doi = {10.1103/PhysRevC.85.064302},
  url = {https://link.aps.org/doi/10.1103/PhysRevC.85.064302}
}

@article{M.Bender_2003_RMP,
  title = {Self-consistent mean-field models for nuclear structure},
  author = {Bender, Michael and Heenen, Paul-Henri and Reinhard, Paul-Gerhard},
  journal = {Rev. Mod. Phys.},
  volume = {75},
  issue = {1},
  pages = {121--180},
  numpages = {0},
  year = {2003},
  month = {Jan},
  publisher = {American Physical Society},
  doi = {10.1103/RevModPhys.75.121},
  url = {https://link.aps.org/doi/10.1103/RevModPhys.75.121}
}

@article{R.Wang_2015_PRC,
  title = {Positioning the neutron drip line and the r-process paths in the nuclear landscape},
  author = {Wang, Rui and Chen, Lie-Wen},
  journal = {Phys. Rev. C},
  volume = {92},
  issue = {3},
  pages = {031303},
  numpages = {5},
  year = {2015},
  month = {Sep},
  publisher = {American Physical Society},
  doi = {10.1103/PhysRevC.92.031303},
  url = {https://link.aps.org/doi/10.1103/PhysRevC.92.031303}
}

@article{G.W.Frederick_1958_PR,
  title = {Superheavy Nuclei},
  author = {Werner, Frederick G. and Wheeler, John A.},
  journal = {Phys. Rev.},
  volume = {109},
  issue = {1},
  pages = {126--144},
  numpages = {0},
  year = {1958},
  month = {Jan},
  publisher = {American Physical Society},
  doi = {10.1103/PhysRev.109.126},
  url = {https://link.aps.org/doi/10.1103/PhysRev.109.126}
}

@article{S.G.Zhou_2003_PRC,
  title = {Spherical relativistic Hartree theory in a Woods-Saxon basis},
  author = {Zhou, Shan-Gui and Meng, Jie and Ring, P.},
  journal = {Phys. Rev. C},
  volume = {68},
  issue = {3},
  pages = {034323},
  numpages = {12},
  year = {2003},
  month = {Sep},
  publisher = {American Physical Society},
  doi = {10.1103/PhysRevC.68.034323},
  url = {https://link.aps.org/doi/10.1103/PhysRevC.68.034323}
}

@article{S.G.Zhou_2010_PRC,
  title = {Neutron halo in deformed nuclei},
  author = {Zhou, Shan-Gui and Meng, Jie and Ring, P. and Zhao, En-Guang},
  journal = {Phys. Rev. C},
  volume = {82},
  issue = {1},
  pages = {011301},
  numpages = {5},
  year = {2010},
  month = {Jul},
  publisher = {American Physical Society},
  doi = {10.1103/PhysRevC.82.011301},
  url = {https://link.aps.org/doi/10.1103/PhysRevC.82.011301}
}

@article{K.Hiroyuki_2013_JPSP,
    author = {Koura ,Hiroyuki and Chiba ,Satoshi},
    title = {Single-Particle Levels of Spherical Nuclei in the Superheavy and Extremely Superheavy Mass Region},
    journal = {Journal of the Physical Society of Japan},
    volume = {82},
    number = {1},
    pages = {014201},
    year = {2013},
    doi = {10.7566/JPSJ.82.014201},
    URL = {https://doi.org/10.7566/JPSJ.82.014201}
}

\end{document}